\documentclass[aps,prb,showpacs,twocolumn,11]{revtex4-2}

\usepackage{graphicx}
\usepackage{amsmath}
\usepackage{amssymb}
\usepackage{bm}
\usepackage{fourier}
\usepackage{baskervald}
\usepackage{pbalance}
\usepackage{hyperref}
\hypersetup{%
	pdftitle={Twisted bilayer graphene as a terahertz plasmonic crystal},%
	pdfauthor={Brian S. Vermilyea and Michael M. Fogler},%
	pdfpagemode={UseNone},%
	pdfstartview={FitH},%
	breaklinks=true,%
	citecolor=blue,%
	colorlinks=true,%
	linkcolor=blue,%
	urlcolor=blue
}

\DeclareMathAlphabet{\mathcal}{OMS}{cmsy}{m}{n}  % mathcal font
\DeclareMathAlphabet{\mathbfcal}{OMS}{cmsy}{b}{n}
\DeclareMathOperator{\re}{Re}
\DeclareMathOperator{\im}{Im}
\DeclareMathOperator{\tr}{Tr}

\begin{document}
\author{Brian S. Vermilyea}
\affiliation{Department of Physics, University of California San Diego, 9500 Gilman Drive, La Jolla, California 92093, USA}

\author{Michael M. Fogler}
\affiliation{Department of Physics, University of California San Diego, 9500 Gilman Drive, La Jolla, California 92093, USA}

\title{Twisted bilayer graphene as a terahertz plasmonic crystal}
\begin{abstract}

We study surface plasmons in minimally-twisted gapped bilayer graphene that contains
a triangular network of partial dislocations (or AB-BA domain walls) hosting topologically protected one-dimensional electronic states.
We show that this system behaves as a plasmonic crystal and we calculate its band structure by solving classical equations of motion for charge dynamics on the network links with impedance boundary conditions at the network nodes.
The plasmon dispersion exhibits several notable features such as multiple gapless branches, flat bands, and dissipationless modes at high-symmetry points.
We compare our network-based formalism with the conventional random phase approximation and discuss when each approach is valid.
Calculations of plasmon waves launched by local scatterers are presented to simulate terahertz nano-imaging experiments.

\end{abstract}
\maketitle

\section{Introduction}
\label{sec:Introduction}

\begin{figure}[t]
	\begin{center}
		\includegraphics[width=2.5in]{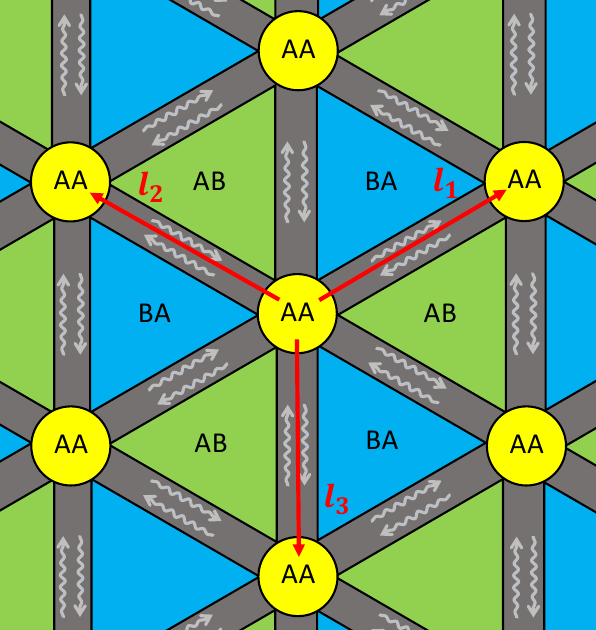}
	\end{center}
	\caption[mTBG structure.]
		{A schematic of the mTBG structure. 
		The AB and BA stacking domains are separated by domain walls (links), which intersect at AA regions (nodes).
		When an interlayer bias is applied, the AB and BA regions are gapped while the domain walls host 1D electron states.
		The wiggly lines represent 1D plasmons that propagate along the links and scatter at the nodes.
		The red arrows indicate the primitive vectors $\mathbf l_1$, $\mathbf l_2$, $\mathbf l_3$ of the moir\'{e} superlattice.}
	\label{fig:network}
\end{figure}

The electronic and lattice structure of twisted bilayer graphene sensitively depend on the twist angle $\theta$. 
While the emergence of superconducting and correlated insulating phases near the ``magic'' angle of $\theta \approx 1.1^{\circ}$ 
has been extensively studied~\cite{Cao2018, Cao2018a},
this material also exhibits interesting physics at smaller $\theta$.
Such a minimally twisted bilayer graphene (mTBG)
undergoes a lattice reconstruction into a triangular (moir\'e) superlattice
of energetically-preferred AB and BA stacking domains separated by narrow domain walls~\cite{Nam2017, Gargiulo2017, Yoo2019},
shown schematically in Fig.~\ref{fig:network}.
For example, a twist angle of $\theta = 0.07^{\circ}$
creates a superlattice with the lattice constant
$L = |\mathbf l_1| = |\mathbf l_2| = |\mathbf l_3| = a_{g} / [2 \sin(\theta / 2)]
\approx 200\, \mathrm{nm}$ (Fig.~\ref{fig:network}), which greatly exceeds the characteristic width of the domain walls $\ell \sim 5-10\,\mathrm{nm}$~\cite{Alden2013}.
In turn, the latter
greatly exceeds the lattice constant $a_{g} = 0.246\,\mathrm{nm}$ of monolayer graphene.

In the experiment, mTBG is often subject to an out-of-plane electric field
due to external gates or charged impurities.
In the presence of such a field, the AB and BA regions develop a bandgap $\Delta_\mathrm{TBG}$,
which can be as high as $250\,\mathrm{meV}$~\cite{Zhang2009},
while AB-BA domain walls remain gapless.
Each domain wall hosts $N_{c} = 2$ bound states whose energy dispersions cross the bandgap. In addition, electrons have the two-fold spin degeneracy $N_s=2$,
yielding the total of $N = N_s N_c = 4$ modes per domain wall per valley.
These one-dimensional (1D) helical states are topologically protected and their propagation directions are opposite for the $\mathrm{K}$ and $\mathrm{K}'$ valleys~\cite{Martin2008, Zhang2013, Vaezi2013, Ju2015, Yin2016, Hasdeo2017}.
These states form a triangular network of conducting channels~\cite{Yoo2019, SanJose2013, Huang2018, Rickhaus2018, Walet2019, Xu2019, Fleischmann2019, Tsim2020, Hou2020}.
The domain walls are the links and their intersections are the nodes (or junctions) of the network, see Fig.~\ref{fig:network}.
It has been theoretically predicted that AB-BA domain walls may host additional non-helical 1D states~\cite{Jiang2017, Moulsdale2024}.
For simplicity, we will not consider this possibility.

Near-field infrared imaging of a doped mTBG~\cite{Jiang2016, Jiang2017, Sunku2018}
revealed that two-dimensional (2D) plasmons in mTBG experience scattering at the domain walls.
This observation is consistent with an enhanced local optical conductivity
due to the topological 1D states.
Direct imaging of these 1D states was not possible 
in those experiments since the Fermi energy was outside the bandgap.

On the theory side, the low-energy electronic band structure of the domain-wall network
has been elegantly described by scattering models~\cite{Efimkin2018, DeBeule2020, Chou2020}.
A basic assumption is that the electron momentum is conserved when they propagate along the links but it can change when they scatter at the nodes. This momentum change can be crudely estimated as the inverse spatial size of the nodes $\sim \ell^{-1}$,
which is small compared to the separation $4\pi / 3 a_g$ of the nearest $\mathrm{K}$ and $\mathrm{K}'$ points in the momentum space.
This implies that scattering at the nodes preserves the valley index.
In particular, electron backscattering at the nodes is negligible in the first approximation.
This property of the mTBG network is rooted in the topological nature of the domain walls
and is not found in previously studied realizations of 1D networks~\cite{Parage1998, Groep2012}.
The cited single-particle network models predict band dispersions $\varepsilon(\mathbf k)$ that are repeated in energy with the period $\Delta \varepsilon = 2\pi \hbar v / L$,
where $v$ is the quasiparticle velocity on the links.
Although realistic electronic structure calculations~\cite{Walet2019, Tsim2020} show more complicated band dispersions, 
such discrepancies can presumably be accounted for by energy dependence of the model parameters.

A conceptually more important omission of these theoretical models is electron-electron interaction.
It is known that low-energy dynamics of interacting 1D electrons is collective~\cite{Solyom1979, Voit1995, Schulz2000, Giamarchi2003, Gogolin2004}.
Therefore, theoretical understanding of the collective modes, such as 1D plasmons, is necessary to describe the response of the system at low frequencies.
Electron-electron interaction effects in AB-BA domain walls and mTBG have been investigated with bosonization methods~\cite{Wieder2015, Wu2019, Lee2021, Chou2021, Hsu2023, Wang2024, Chang2025}
but plasmon modes were not addressed in these studies.

In this paper, we demonstrate that mTBG is a special type of a plasmonic crystal.
Its 2D plasmon modes are built out of 1D plasmons that propagate along the AB-BA links
and are scattered at AA nodes.
The plasmon band structure exhibits unique features including multiple gapless branches,
flat bands, and dissipationless modes at high-symmetry points. 
For typical parameters the frequencies of mTBG plasmons
lie in the terahertz (THz) domain, which may be of interest for nanophotonics applications.

The rest of this paper is organized as follows.
In Sec.~\ref{sec:Introduction2}, we introduce a physical picture and two
theoretical approaches we employ to study the plasmons.
In Sec.~\ref{sec:RPA}, we analyze the problem within the standard random phase approximation (RPA).
In Sec.~\ref{sec:scattering}, we lay a groundwork for
an alternative approach, which we refer to as the
plasmon network model,
by treating the plasmon scattering problem at a single junction.
In Sec.~\ref{sec:PNM}, we
compute the plasmonic spectrum within the PNM and compare the results with those from the RPA.
In Sec.~\ref{sec:imaging},
we simulate signatures of mTBG plasmons in near-field imaging.
In Sec.~\ref{sec:discussion}, we conclude the paper with a brief summary and an outlook.
Additional technical details are provided in the Appendix.

%%%%%%%%%%%%%%%%%%%%%%%%%%%%%%%%%%%%%%%%%%%%%%%%%%%%%%%%%%%%%%%%%%%%%%%%%%%%%%
\section{Qualitative discussion} \label{sec:Introduction2}

\begin{figure}[b]
	\begin{center}
		\includegraphics[width=2.75in]{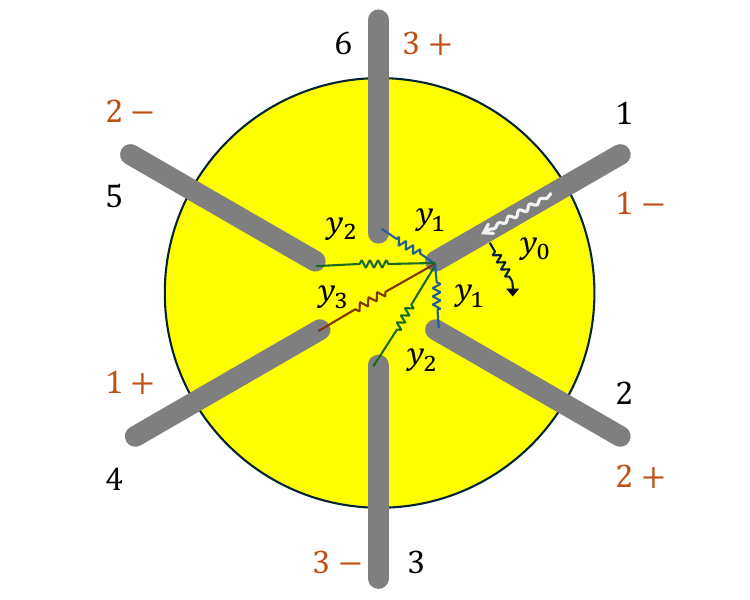}
	\end{center}
	\caption[Effective circuit]
	{Effective circuit for a junction of six leads (the thick solid lines). The numbers indicate two different schemes of labeling the leads. For clarity, only the connections originating from lead $1$ ($1-$) are shown, e.g., admittance connecting it to the ground is $y_0$, the ones wiring it to the nearest leads are $y_1$, \textit{etc}. The remaining leads are wired the same way.}
	\label{fig:junction_circuit}
\end{figure}

Rather than attempting to construct a general theory of mTBG plasmons,
we consider two limiting cases distinguished by the relation between the network lattice constant $L$ and the single-particle phase coherence length $L_\varphi$~\cite{Datta1995}.
When $L_\varphi\gg L$, the electrons maintain phase coherence upon traversing the links.
In this regime, non-Fermi-liquid effects may cause some renormalization of the bare scattering parameters but they do not qualitatively change the response of the system.
Therefore, we employ the conventional RPA to calculate the dielectric function and the plasmon dispersion.
In general, the RPA becomes more accurate with increasing space dimension $d$ and number of modes $N$.
Applied to mTBG, where $d = 2$ and $N = 4$, it may be a reasonable approximation in the coherent regime.

In the opposite limit, $L_\varphi\ll L$,
phases of single-particle states are randomized while they propagate along the links, due to inelastic scattering.
The physically relevant degrees of freedom are truly collective: the total charge density and the total current.
Their dynamics is governed by the Tomonaga-Luttinger liquid (TLL) theory~\cite{Voit1995, Schulz2000, Giamarchi2003, Gogolin2004}. 
Each time a plasmon is scattered at a node,
additional bosonic degrees of freedom (electron-hole pairs) are excited
but they quickly relax and dissipate some energy.
Assuming no other momentum non-conserving processes (like phonon scattering)
are present, the plasmons propagate ballistically from one node to the next.

Both $L_\varphi < L$ and $L_\varphi > L$ cases can be realized
by varying control parameters such as temperature $T$ and frequency $\omega$.
In experiment, the dephasing length
of mTBG was observed to vary inversely proportional to temperature,
\begin{equation}
	L_\varphi \propto \frac{1}{T}\,,
	\label{eqn:L_phi_exp}
\end{equation}
in the range from $T_\mathrm{min} \sim 10\,\mathrm{K}$ to $T_\mathrm{max} = 120\,\mathrm{K}$~\cite{Xu2019}.
From Fig.~2 of Ref.~\onlinecite{Xu2019} we estimated $L_\varphi \approx 80\,\mathrm{nm}$ at $T = T_\mathrm{max}$, which is shorter than $L \approx 140\,\mathrm{nm}$ in the studied sample.
On the other hand, $L_\varphi \gg L$ at $T = T_\mathrm{min}$.
These results are consistent with the theory of dephasing in a TLL~\cite{LeHur2005, Gornyi2005, LeHur2006}, which predicts
$L_\varphi \propto \hbar v/T$ with a proportionality constant of order unity that depends on the electron-electron interaction strength.
Similarly, at low temperature but finite frequency such that $\omega \gg T / \hbar$ we expect $L_\varphi \propto v/\omega$.

Since Coulomb interaction mixes plasmons of all propagation directions, plasmon backscattering at the nodes can be significant
even though the suppressed intervalley scattering makes it negligible for single electrons.
The technical term used to describe it in the context of 1D systems
is the $g_2$-coupling~\cite{Solyom1979, Voit1995, Schulz2000, Giamarchi2003, Gogolin2004}.
Taking into account such scattering channels is a key distinction of our theory from a prior work~\cite{Brey2020} where the plasmon backscattering was discarded.

In our plasmon network model (PNM), the plasmon scattering at a junction can be visualized in terms
of an effective circuit, Fig.~\ref{fig:junction_circuit},
where interlink admittances $y_j$ relate currents and voltage differences between the links.
These parameters can be organized into a $6 \times 6$ admittance matrix $\mathbf{Y}$.
Such an approach is customary in the theory of microwave transmission lines~\cite{Montgomery1987pom}.
In the actual calculation, we work not with $y_j$'s but with the eigenvalues $Y_m$ of matrix $\mathbf{Y}$,
which are directly related to the plasmon scattering phase shifts in different angular momentum channels.
We show that these phase shifts depend on the strength of electron-electron interactions.
The latter is expressed in terms of the TLL parameter $K$,
where $K = 1$ for non-interacting electrons
 and $K < 1$ for repulsive interactions~\cite{Voit1995, Schulz2000, Giamarchi2003, Gogolin2004}.
In the weakly interacting limit, $K \to 1$, the admittance matrix is determined by the single-particle scattering probabilities
in accordance with the Landauer formula~\cite{Kamenev2001}.
The eigenvalues $Y_m$ are purely real and positive;
hence, the junction behaves as a circuit of resistors.
Coulomb interactions among the links generate imaginary corrections to the phase shifts.
This is equivalent to adding some capacitors in parallel to the resistors.

As mentioned above, many-body interaction renormalizes the bare scattering amplitudes,
see Refs.~\onlinecite{Kane1992, Furusaki1993, Matveev1993, Matveev1993a} for the early work and
Refs.~\onlinecite{Chamon2003, Wieder2015, Wu2019, Lee2021, Chou2021, Hsu2023, Wang2024, Chang2025} for generalizations and applications to AB-BA domain walls.
This effect may produce further real and imaginary contributions to the admittance
of the junctions (similar to those found in 1D systems~\cite{Guinea1995, Houzet2024}).
These corrections have a power-law dependence on $T$ and $\omega$ with an exponent that scales roughly as $1 / N$~\cite{Matveev1993a}.
In mTBG, where $N = 4$, we expect these corrections to be small and neglect them for simplicity.

As we show below, the RPA and the PNM give plasmon spectra that agree in their overall structure. However, the RPA predicts more fine features compared to the PNM.
In both theories these spectra are not strictly periodic in frequency,
unlike the single-particle ones~\cite{Efimkin2018, DeBeule2020, Chou2020},
because the plasmon velocity on the links is momentum-dependent even without scattering.
Within the RPA, the plasmon modes are punctuated by numerous narrow particle-hole continua.
The plasmons remain sharp outside these continua but experience Landau damping inside of them.
Within the PNM, the damping has a more uniform character in absolute units,
so that in relative terms it has a stronger effect as frequency decreases.
In the asymptotic limit $\omega \to 0$,
all the plasmon modes are overdamped according to the PNM
whereas they remain sharp within the RPA.

\section{Random phase approximation} \label{sec:RPA}

We begin this section with a theoretical model for the single-particle states of the domain-wall network of mTBG.
Next, we explain our choice for the electron-electron interaction model.
Using these ingredients, we compute the dielectric function within the RPA.
Finally, we discuss the predictions of the RPA for the plasmon dispersion of mTBG.

\subsection{Single-particle spectrum of mTBG}
\label{sub:single-particle_mTBG}

\begin{figure}[t]
	\begin{center}
		\includegraphics[width=2.75in]{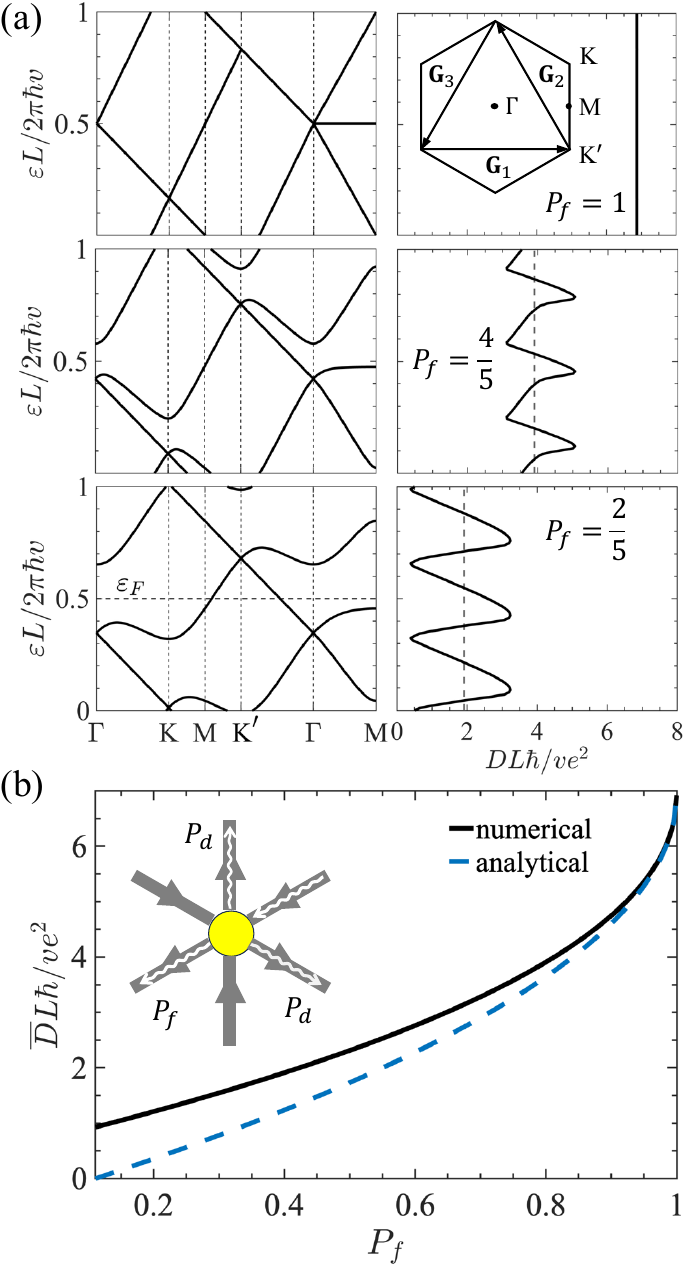}
	\end{center}
	\caption[Band structure and Drude weight]{
		(a) Band structure along high-symmetry lines in the the moir\'e Brillouin zone (left) 
		and corresponding Drude weight [Eq.~\mbox{\eqref{eqn:drude}}] \textit{vs}. Fermi energy (right)
		for $P_f = 1$ (top), $P_f = 0.8$ (middle), and $P_f = 0.4$ (bottom).
		Only the three bands of the $\mathrm{K}$ valley are shown.
		The vertical dashed lines indicate the average Drude weight $\overline D$.
		The horizontal dashed line in the bottom left panel denotes the Fermi energy $\varepsilon_F$ used in the calculation of Figs.~\ref{fig:bands} and \ref{fig:bands2}.
		Inset at top right: the moir\'e Brillouin zone. Arrows indicate the reciprocal lattice vectors defined in Eq.~\eqref{eqn:G_vectors}.
		(b) Average Drude weight $\overline D$ as a function of $P_f$.
		The solid line is Eq.~\eqref{eqn:average_Drude} evaluated numerically,
		and the dashed line is Eq.~\eqref{eqn:D_avg}.
		Inset: diagram showing allowed single-particle scattering directions.
	}
	\label{fig:drude_weight}
\end{figure}

To obtain the single-particle band structure we employ the network model of Ref.~\onlinecite{Efimkin2018}.
This model has a single parameter: the forward scattering probability $P_f$.
As explained in Sec.~\ref{sec:Introduction}, the backscattering probability is zero.
The allowed scattering processes at a junction are shown
in the inset of Fig.~\ref{fig:drude_weight}(b).
In addition to forward scattering, they include deflections
by angles $\pm 2\pi / 3$ with probability $P_d = (1 - P_f) / 2$.
Effectively, each of the three intersecting domain wall acts as a partially transparent beam-splitting mirror.
As also mentioned in Sec.~\ref{sec:Introduction},
each domain wall hosts $N = 4$ 1D helical modes.
For simplicity, we neglect scattering between these modes
although the model can be generalized to include that~\cite{DeBeule2020}.

The single-particle scattering matrix of a junction takes the form
\begin{equation}
	\mathbf{S}_0 =
	\begin{pmatrix}
		0 & s_d & 0 & s_f & 0 & s_d\\
		s_d & 0 & s_d & 0 & s_f & 0\\
		0 & s_d & 0 & s_d & 0 & s_f\\
		s_f & 0 & s_d & 0 & s_d & 0\\
		0 & s_f & 0 & s_d & 0 & s_d\\
		s_d & 0 & s_f & 0 & s_d & 0
	\end{pmatrix}
.
\label{eqn:S_0}
\end{equation}
The six rows (columns) of $\mathbf{S}_0$ correspond to the outgoing (incoming) states of six leads numbered $1$--$6$ in Fig.~\ref{fig:junction_circuit}.
Note that the scattering amplitudes $s_f$ ($s_d$) are equal, up to irrelevant phases, to the diagonal (off-diagonal) elements of the scattering matrix defined in Eq.~(9) of Ref.~\onlinecite{Efimkin2018}.
Without loss of generality, we can choose the deflection amplitude
$s_d = i \sqrt{P_d}$ to be imaginary.
The unitarity of $\mathbf{S}_0$ then determines
$s_f = -\tfrac12\left(\sqrt{4P_f-P_d}+i\sqrt{P_d}\right)$.
The eigenvalues of $\mathbf{S}_0$ have the form $e^{2 i \delta_{0, m}}$ where scattering phase shifts $\delta_{0, m}$ are real.
We index these phase shifts by $m = 0, \pm 1, \pm 2, 3$, which is the angular momentum modulo $6$.
The corresponding eigenvectors $\mathbf{z}_m$ have components
\begin{equation}
{z}_{m, a} = \frac{1}{\sqrt{6}}\, e^{i \pi m a / 3} ,
\qquad a = 0, 1, \ldots, 5\,.
\label{eqn:z_m}                   
\end{equation}
Diagonalizing $\mathbf{S}_0$, we find the phase shifts as follows:
\begin{subequations}
	\begin{align}
		\delta_{0, 0} &= \frac{\pi}{2} - \frac12 \arcsin\left(\frac32 \sqrt{P_d}\, \right)\,,
		\label{eqn:delta_0}\\
		\delta_{0, \pm 1} &= \frac{\pi}{2} - \delta_{0, 0}\,,
		\label{eqn:delta_1}\\
		\delta_{0, \pm 2} &= \pi - \delta_{0, 0} \,,
		\label{eqn:delta_2}\\
		\delta_{0, 3} &= \frac{\pi}{2} + \delta_{0, 0}\,.
		\label{eqn:delta_3}
	\end{align}
	\label{eqn:S_eigenvalues}
\end{subequations}
Note that
\begin{equation}
                   P_d = \frac49 \sin^2 2 \delta_{0, 0} \,,
\label{eqn:P_d_from_delta}                   
\end{equation}
so the admissible values of $P_d$ are in the range $0 \leq P_d \leq 4 / 9$.
According to some studies~\cite{Qiao2014,Anglin2017,You2022}, deflection is favored over forward scattering, in which case $1 / 3 < P_d \leq 4 / 9$.

Following Ref.~\onlinecite{Efimkin2018}, let us review the computation of the single-particle states.
Let us choose a unit cell of the network such that it is centered at a node and
contains three links of length $L$. One possible way to do so is to start with the usual hexagonal Wigner-Seitz cell $-\pi < \mathbf{G}_{a} \cdot \mathbf{r} < \pi$,
where $\mathbf{G}_{a}$ are three shortest reciprocal lattice vectors:
\begin{equation}
\mathbf{G}_a = \frac{4 \pi}{3 L} \left(\,
\hat{\mathbf{l}}_a - \hat{\mathbf{l}}_{a + 1}\right),
\quad
a = 1, 2, 3\: (\mathrm{mod}\ 3) ,
\label{eqn:G_vectors}
\end{equation}
and $\hat{\mathbf{l}}_a = \mathbf{l}_a / L$ is the unit vector in the direction of the lattice vector $\mathbf{l}_a$ (Fig.~\ref{fig:network}).
This Wigner-Seitz cell contains six links of \emph{half} length, so we add the remaining halves of the links parallel to the lattice vectors $\mathbf{l}_a$ and remove the half-links anti-parallel to $\mathbf{l}_a$.
Each link of such a unit cell has two propagation directions $(+)$ and $(-)$, 
so in total there are six independent wavefunction amplitudes per unit cell per mode.
Going clockwise from link $\mathbf{l}_1$ (Fig.~\ref{fig:junction_circuit}), we organize these amplitudes into a six-component column vector
\begin{equation}
	{\bm{\Psi}}
	= \left(\psi_{1-}, \psi_{2+}, \psi_{3-}, \psi_{1+}, \psi_{2-}, \psi_{3+}\right)^T .
\end{equation}
The even (odd) components represent the states that belong to the $\mathrm{K}'$ ($\mathrm{K}$) valley.
Consider an eigenstate with a crystal momentum vector $\mathbf{k}$ that belongs to the moir\'e Brillouin zone $-\pi / L < k_a < \pi / L$,
where $k_a = \mathbf{k} \cdot \hat{\mathbf{l}}_a$.
At energy $\varepsilon$, each component of ${\bm{\Psi}}$ accumulates a phase $\varepsilon L / \hbar v$
traversing the length of the link.
Adding the appropriate Bloch phase factors to ${\bm{\Psi}}$,
we obtain the vectors of incoming and outgoing amplitudes at a node:
\begin{equation}
	\bm{\Psi}_\mathrm{in} = e^{i \varepsilon L / \hbar v}
	\left(\begin{array}{l}
		\psi_{1-} \\ \psi_{2+} e^{-i k_2 L} \\ \psi_{3-} \\ 
		\psi_{1+} e^{-i k_1 L} \\ \psi_{2-} \\
		\psi_{3+} e^{-i k_3 L}
	\end{array}\right), \quad
	\bm{\Psi}_\mathrm{out} = \left(\begin{array}{l}
	   \psi_{1+} \\ \psi_{2-} e^{-i k_2 L} \\ \psi_{3+} \\ 
		\psi_{1-} e^{-i k_1 L} \\ \psi_{2+} \\
		\psi_{3-} e^{-i k_3 L}
	\end{array}\right) .
\end{equation}
The amplitudes $\bm{\Psi}_\mathrm{in}$ and $\bm{\Psi}_\mathrm{out}$ are related via the scattering matrix $\mathbf{S}_0$:
$\bm{\Psi}_\mathrm{out} = \mathbf{S}_0 \, \bm{\Psi}_\mathrm{in}$.
Therefore, the single-particle states are obtained by solving the eigenproblem
\begin{equation}
	\mathbf{T}_0\, {\bm{\Psi}} = e^{-i \varepsilon L / \hbar v}\, {\bm{\Psi}} ,
	\label{eqn:single-particle_eigenproblem}
\end{equation}
where
\begin{gather}
	\mathbf{T}_0 =
	\begin{pmatrix}
		0 & \mathbf{I}_3 \\ \mathbf{I}_3 & 0
	\end{pmatrix}
	\bm{\Lambda}_0^{-1} \mathbf{S}_0\, \bm{\Lambda}_0 ,
	\\
	\bm{\Lambda}_0 = \operatorname{diag} \left(1, e^{-i k_2 L}, 1, e^{-i k_1 L}, 1, e^{-i k_3 L}\right) ,
\end{gather}
and $\mathbf{I}_3$ is the $3 \times 3$ identity matrix.
The chosen scattering matrix $\mathbf{S}_0$ does not mix the valleys. As a result, our $6$-component problem separates into two independent $3$-component ones,
\begin{equation}
	\mathbf{T}^{\nu}_0\, {\bm{\Psi}^{\nu}} = e^{-i \varepsilon L / \hbar v}\, {\bm{\Psi}^{\nu}} ,
	\qquad {\nu} = + \text{ or } -\,,
	\label{eqn:single-particle_eigenproblem_valley}
\end{equation}
where, e.g.,  ${\bm{\Psi}}^{+} = \left(\psi_{2+}, \psi_{1+}, \psi_{3+}\right)^T$ and
\begin{equation}
	\mathbf{T}^{+}_0
	= \begin{pmatrix}
		s_f & s_d & s_d\\
		s_d & s_f & s_d\\
		s_d & s_d & s_f
	\end{pmatrix} 
	\left(\begin{array}{lll}
	e^{-i k_2 L} & 0 & 0\\
	0 & e^{-i k_1 L} & 0\\
	0 & 0 & e^{-i k_3 L}
\end{array}\right) .
\label{eqn:T_valley}
\end{equation}
To find the electronic spectrum, we numerically solve the eigenvalue problem Eqs.~{\eqref{eqn:single-particle_eigenproblem_valley}} and {\eqref{eqn:T_valley}}. 
The eigenvectors give the wavefunction amplitudes on each link.
The phase of the eigenvalue $e^{-i \varepsilon L / \hbar v}$ determines the band energies.

In Fig.~\ref{fig:drude_weight}(a,left), we plot the band structure computed for a few values of $P_f$.
Only the three bands of the $\mathrm{K}$ valley (i.e., $\nu = +$ bands) are shown; those corresponding to the $\mathrm{K}'$ valley ($\nu = -$) can be obtained by reversing the crystal momentum.
Also, we plot only one period $\varepsilon_L=2\pi \hbar v / L$ in energy.
The top plot in Fig.~\ref{fig:drude_weight}(a) corresponds to $P_f = 1$, where $\varepsilon_l(\mathbf{k}) = \hbar v (k_l + \pi / L)$,
so that the surface $\varepsilon = \varepsilon_l(\mathbf k)$ in the energy-momentum space consists of three intersecting planes.
This is why the plot is composed of straight lines.
Along the high-symmetry lines connecting $\Gamma$, $\mathrm{K}$, and $\mathrm{K}'$,
two-fold band degeneracies (not counting the valley) occur because $k_l = k_{l'}$ for some $l$ and $l'$.
For $P_f$ slightly less than unity,
the off-diagonal terms in $\mathbf{T}^{+}_0$ lift these degeneracies.
In particular, near the $k_l = k_{l'}$ lines but away from $\Gamma$, $\mathrm{K}$, and $\mathrm{K}'$,
the band energies (modulo $2\pi \hbar v / L$) are given by
\begin{equation}
	\varepsilon_{l l',\pm}(\mathbf{k}) \simeq
	{\hbar v}
	\left[\frac{\pi}{L}  - \frac{\sqrt{P_d}}{2 L} + \frac{k_l + k_{l'}}{2} \pm \sqrt{\left(\frac{k_l - k_{l'}}{2}\right)^2 + \frac{P_d}{L^2}}\, \right] ,
	\label{eqn:single-particle_bands_small_deflection}
\end{equation}
so that $\varepsilon_{l l', +}(\mathbf{k}) - \varepsilon_{l l', -}(\mathbf{k}) \simeq 2 \hbar v \sqrt{P_d} / L$
along the aforementioned high-symmetry lines.
This approximately constant energy splitting is apparent for the bands with a negative slope in the middle row of Fig.~\ref{fig:drude_weight}(a) where $P_d = 0.1$.
As the $\Gamma$ point is approached, the two energies $\varepsilon_{l l',\pm}(\mathbf{k})$ are perturbed by the proximity of the third band.
Precisely at the $\Gamma$ point,
there is a triple degeneracy if $P_d = 0$ [the top row in Fig.~\ref{fig:drude_weight}(a)],
which is split into a singlet of energy $2 \hbar v (\pi - \delta_{0, 0}) / L$ and a doublet (a Dirac point) of energy $2 \hbar v \delta_{0, 0} / L$ if $P_d > 0$,
see the middle and the bottom rows of Fig.~\ref{fig:drude_weight}(a).
The same pattern is repeated near $\mathrm{K}$ and $\mathrm{K}'$ points.

%%%%%%%%%%%%%%%%%%%%%%%%%%%%%%%%%%%%%%%%%%%%%%%%%%%%%%%%%%%%%%%%%%%%%%%%
\subsection{2D and 1D interaction kernels}
\label{sub:kernels}

When computing the electrodynamic response of mTBG,
we envision that the electron motion in the direction perpendicular to the domain walls is characterized by a certain fixed density profile.
The electron density per unit area $n_{\mathrm{2D}}(\mathbf{r})$ at an arbitrary position $\mathbf{r}$ is the convolution of this profile with a 1D electron density $n_{a, \mathbf{R}}(x)$.
The latter has the units of inverse length and is defined at points $\mathbf{r}$
that belong to the network of zero-thickness links,
$\mathbf{r} = \mathbf{R} + \hat{\mathbf{l}}_a x$.
Here $0 < x < L$ is the distance from $\mathbf{r}$ to the center $\mathbf{R} = -m_3 \mathbf{l}_1 + m_2 \mathbf{l}_2$ of a nearby unit cell,
$m_a$'s are the integer parts of $\mathbf{G}_a \cdot \mathbf{r} / (2\pi)$.

For calculation of $n_{\mathrm{2D}}(\mathbf{r})$ and other convolutions we may need, we take advantage of the Fourier transform, which we denote by a tilde, e.g.,
$\tilde{n}_{\mathrm{2D}}(\mathbf{q})$.
Whenever the argument of the Fourier transform is
a vector, as is $\mathbf{q}$ in this example,  
the transform is 2D; if its argument is a scalar, then it is a 1D Fourier transform.
All the quantities also depend on time through a common $e^{-i \omega t}$ factor,
which we omit.

A charge excitation with a given crystal momentum $\mathbf{q}$ creates a density 
perturbation that can be expanded in either 1D or 2D Fourier series:
\begin{align}
	n_{a,\mathbf{R}}(x) &= e^{i\mathbf{q} \cdot \mathbf{r}}
	\sum_{p = -\infty}^\infty n_{a p} e^{2\pi i p x / L}, \quad \mathbf{r} = \mathbf{R} + \hat{\mathbf{l}}_a x,
\label{eqn:n_1D}\\
	n_{\mathrm{2D}}(\mathbf{r}) &= \frac{1}{L_{\bot}} \sum_{\mathbf{G}}
e^{i (\mathbf{q} + \mathbf{G}) \cdot \mathbf{r}} \,
\sum_{p} X_{\mathbf{G}, a p}
\rho_{\bot}\left[ (\mathbf{q} + \mathbf{G}) \cdot \hat{\mathbf{l}}_{a \bot} \right]
 n_{a p}\,,
\label{eqn:n_2D}\\
X_{\mathbf{G}, a p} &\equiv \int\limits_0^L \frac{d x}{L}\, e^{-2\pi i p x / L}
 e^{i \mathbf{G} \cdot \hat{\mathbf{l}}_a x} = \delta_{\!p, g_a} ,
\quad
g_a = \frac{\mathbf{G} \cdot \mathbf{l}_a}{2\pi}\, .	
\label{eqn:X}
\end{align}
Here $L_{\bot} = {\sqrt{3}\, L} / {2}$ is the separation between nearest parallel domain walls (Fig.~\ref{fig:network}),
$\mathbf{G}$ are reciprocal lattice vectors,
and $\hat{\mathbf{l}}_{a \bot} = \hat{\mathbf{z}} \times \hat{\mathbf{l}}_{a}$
is the unit vector perpendicular to the $a\,$th link.
Function $\rho_{\bot}(k)$ in Eq.~\eqref{eqn:n_2D} is a form-factor,
i.e., the 1D Fourier transform of the domain wall density profile mentioned above.
It is normalized by the condition $\rho_{\bot}(0) = 1$ and becomes small at $k \gg 1 / \ell$.

The density perturbation given by Eq.~\eqref{eqn:n_2D} induces an energy shift $e \Phi_{\mathrm{ind}}(\mathbf{r})$ for electrons on the links,
which is equal to the convolution of the density $n_{\mathrm{2D}}(\mathbf{r})$ with the
electron interaction kernel $U_{\mathrm{2D}}(\mathbf{r})$ and the domain-wall density profile.
Here $e = -|e|$ is the electron charge.
Utilizing Eq.~\eqref{eqn:n_2D},
we can express this periodically varying potential shift as another Fourier series
\begin{align}
e \Phi_{\mathrm{ind}}(\mathbf{r}) =& \frac{1}{L_{\bot}} \sum_{\mathbf{G}}
e^{i (\mathbf{q} + \mathbf{G}) \cdot \mathbf{r}} \tilde{U}_{\mathrm{2D}}(\mathbf{q} + \mathbf{G})
\nonumber\\
&\times 
\sum_{ap} X_{\mathbf{G}, ap}
\rho_{\bot}\left[ (\mathbf{q} + \mathbf{G}) \cdot \hat{\mathbf{l}}_{a \bot} \right]
 n_{ap}
\label{eqn:Phi_from_n_bare}\\
\approx& \frac{1}{L_{\bot}} \sum_{\mathbf{G}}
e^{i (\mathbf{q} + \mathbf{G}) \cdot \mathbf{r}} \tilde{U}(\mathbf{q} + \mathbf{G})
\sum_{ap} X_{\mathbf{G}, ap} n_{ap} .
\label{eqn:Phi_from_n}
\end{align}
The form-factors $\rho_{\bot}$ in Eq.~\eqref{eqn:Phi_from_n_bare} ensure that the series converges at $|\mathbf{G}| \sim 1 / \ell$
but make the formulas unnecessarily complicated.
In Eq.~\eqref{eqn:Phi_from_n} we drop these factors and
achieve the convergence by replacing $\tilde{U}_{\mathrm{2D}}(\mathbf{q})$
with a regularized kernel $\tilde{U}(\mathbf{q})$, which corresponds to
a real-space interaction potential $U(\mathbf{r})$ well-behaved at short distances $|\mathbf{r}| \alt \ell$.

In this paper, we consider two model forms of such $U(\mathbf{r})$.
One is the bare Coulomb interaction,
\begin{equation}
	U(\mathbf{r}) = \frac{e^2}{\kappa}\frac{1}{|\mathbf{r}| + \ell}\,,
\label{eqn:bare_kernel_2D}
\end{equation}
where $\kappa$ is the dielectric constant.
The other is a screened interaction
\begin{align}
	U(\mathbf{r}) &= \frac{e^2}{\kappa}\,\frac{1}{|\mathbf{r}| + \ell}
	- \frac{e^2}{\kappa}\,
	\frac{1}{\sqrt{|\mathbf{r}|^2 + 4 d_g^2}}\,,
	\label{eqn:screened_kernel_2D}
\end{align}
which includes the image term due to a conducting plate
separated by a distance $d_g$ from the sample.

For working with the 1D Fourier basis we will need the matrix elements
\begin{align}
\begin{split}
\tilde U_{ap,a'p'}(\mathbf q) =& \frac{1}{L}\, \sum_{\mathbf R}
\int\limits_0^L d x e^{-i q_{a p} x} \int\limits_0^L d x'e^{i q_{a' p'} x'}
\\
&\times e^{-i\mathbf{q} \cdot \mathbf{R}}
U\left(\mathbf{R} + \hat{\mathbf l}_a x - \hat{\mathbf l}_{a'} x'\right) 
\label{eqn:U1}
\end{split}\\
\mbox{} =& \frac{1}{L_{\bot}}\, \sum_{\mathbf{G}}
\tilde{U}\left(\mathbf{q} + \mathbf{G}\right)
X_{\mathbf{G}, a p} X_{\mathbf{G}, a' p'} \,,
\label{eqn:U1_G}
\end{align}
where we use a convenient short-hand notation
\begin{equation} 
	q_{a p}	= q_a + \frac{2\pi p}{L} \,.
\label{eqn:q_ap}
\end{equation}
For off-diagonal elements, $a \neq a'$, the last sum reduces to a single term:
\begin{align} 
	\tilde U_{ap, a' p'}(\mathbf q) &=
	\frac{1}{L_{\bot}}\, \tilde{U}\left(\mathbf{q} + \frac{p}{g}\, \mathbf{G}_{a' + 1} + \frac{p'}{g'}\, \mathbf{G}_{a + 1} \right)\,,
\label{eqn:U_aa'}\\
g &= \frac{1}{2\pi}\, \mathbf{G}_{a' + 1} \cdot \mathbf{l}_{a}\,,
\quad 
g' = \frac{1}{2\pi}\, \mathbf{G}_{a + 1} \cdot \mathbf{l}_{a'}\,.
\end{align}
If the interaction range $d_g$ is short enough, such that $d_g \ll L$ and $L |\mathbf{q}|, p, p' \ll L / d_{g}$,
then these matrix elements are relatively small and approximately equal to one another:
\begin{equation} 
	\tilde U_{ap, a' p'}(\mathbf q)
	\simeq
	\frac{1}{L_{\bot}}\, \tilde{U}\left(\mathbf{0}\right)
	\simeq 
	\frac{8\pi}{\sqrt{3}}\, \frac{e^2}{\kappa}\, \frac{d_g}{L}
	\,,\quad a \neq a'\,.
	\label{eqn:U_aa'_screened}
\end{equation}
The diagonal elements are given by
\begin{equation} 
	\tilde U_{ap, a p'}(\mathbf q) \simeq \tilde{U}(0)\delta_{p p'}
	\simeq \frac{2 e^2}{\kappa}\, \ln \left(\frac{d_g}{\ell}\right) \delta_{p p'}\,,
\label{eqn:U_0}
\end{equation}
see Eq.~\eqref{eqn:Fourier_screened_kernel} below.
These formulas can also be derived from the real-space integral in Eq.~\eqref{eqn:U1},
which is dominated by small regions near the junction of the links, $0 < x, x' \alt d_g$,
$0 < L - x, x' \alt d_g$, and $0 < x, L - x' \alt d_g$,
where all the exponential factors in the integrand can be dropped,
see Appendix~\ref{app:scattering}.

Finally, one more definition is the 1D Fourier transform of $U(x)$:
\begin{align}
	\tilde{U}(q) &\simeq \frac{2 e^2}{\kappa}\,
	\ln \left(\frac{d_g}{\ell}\right),
	\quad  q \ll d_g^{-1}
	\label{eqn:Fourier_screened_kernel}\\
	& \simeq \frac{2 e^2}{\kappa}\, \ln \left(\frac{A}{q\ell}\right)\,,
	\quad  d_g^{-1} \ll q \ll \ell^{-1}
	\label{eqn:Fourier_kernel}
\end{align}
with $A = 0.561$.
As will become clear in the following, for $d_g$ much smaller than the plasmon wavelength
we can use Eq.~\eqref{eqn:Fourier_screened_kernel} at all $q$, i.e.,
\begin{equation}
	\tilde{U}(q) = \frac{2 e^2}{\kappa}\, \ln \left(\frac{d_g}{\ell}\right) = \mathrm{const}\,.
	\label{eqn:short_range_kernel}
\end{equation}
This is equivalent to setting $U(x) = \tilde{U}(0) \delta(x)$ on the links,
which we refer to as the short-range interaction model.

%%%%%%%%%%%%%%%%%%%%%%%%%%%%%%%%%%%%%%%%%%%%%%%%%%%%%%%%%%%%%%%%%%%%%%%%
\subsection{RPA and the plasmon spectrum in mTBG}
\label{sub:RPA_mTBG}

Within the RPA, the electrons respond to the mean-field energy shift
$e \Phi(\mathbf{r}) =  e \Phi_{\mathrm{ext}}(\mathbf{r}) + e \Phi_{\mathrm{ind}}(\mathbf{r})$
where $\Phi_{\mathrm{ext}}(\mathbf{r})$ is an external potential and
$\Phi_{\mathrm{ind}}(\mathbf{r})$ is the induced potential
[Eq.~\eqref{eqn:Phi_from_n}].
Similar to the electron density, $e \Phi(\mathbf{r})$ can be expanded in a 1D Fourier series:
\begin{align}
	e \Phi(\mathbf{r})
	&= e^{i\mathbf{q} \cdot \mathbf{r}}
	\sum_{p} e \Phi_{a p} e^{2\pi i p x / L} ,
	\quad \mathbf{r} = \mathbf{R} + \hat{\mathbf{l}}_a x\,.
\label{eqn:Phi_1D}
\end{align}
The induced electron density is related to $e \Phi(\mathbf{r})$ via
\begin{equation}
n_{ap} = \sum_{a'p'} P^{(0)}_{ap, a'p'}(\mathbf{q}, \omega)\, e\Phi_{a'p'},
\end{equation}
where $P^{(0)}_{ap,a'p'}(\mathbf{q}, \omega)$ is the free-particle polarization function:
\begin{align}
	P^{(0)}_{ap, a'p'} =& P^{+}_{ap, a'p'}
	+ P^{-}_{ap, a'p'}\,,
\label{eqn:polarization_as_sum}\\
\begin{split}
	P^{\nu}_{ap, a'p'}(\mathbf{q}, \omega) =& N L_{\bot} \sum_{l l'}
	 \int \frac{d^2 k}{(2\pi)^2}
	 \rho_{l l', ap} \rho_{l l', a'p'}^*
	 \\ 
	\mbox{} &\times \frac{f\left[\varepsilon_{l}(\mathbf{k})\right]
		 - f\left[\varepsilon_{l'}(\mathbf{k} + \nu \mathbf{q})\right]}
	                 {\hbar(\omega + i\gamma_d) + \varepsilon_{l}(\mathbf{k})
	                 	   - \varepsilon_{l'}(\mathbf{k} + \nu \mathbf{q})},
\end{split}
\label{eqn:polarization2}\\
\begin{split}
	\rho_{l l', ap} =& \int \limits_0^L d x\, e^{-i q_{a p} x} \Psi_{l,a}^*(\mathbf k)
	\Psi_{l',a}(\mathbf{k} + \nu \mathbf{q}),
\end{split}
\label{eqn:rho2}\\
	\Psi_{l,a}(\mathbf k) =& \frac{1}{\sqrt{L}} \psi_{l,a}(\mathbf k)
	 e^{i \varepsilon_{l}(\mathbf{k}) x / \hbar v}\,.
\label{eqn:Psi2}
\end{align} 
Here $\nu = \pm$ is again the valley index,
the $\mathbf{k}$-integration is over the moir\'e Brillouin zone,
$f(\varepsilon) = \Theta(\varepsilon_F - \varepsilon)$ is the Fermi distribution function,
$\varepsilon_{l}(\mathbf{k})$ is the energy of the $l\,$th band in valley $\nu = +$,
$\psi_{l,a}(\mathbf{k})$ is the corresponding wavefunction amplitude on link $a$,
normalized such that $\sum_a |\psi_{l,a}(\mathbf{k})|^2 = 1$,
and $\gamma_d$ is a small damping parameter, added for regularization purposes.

In general, functions $P^{\mu}_{a''p'',a'p'}$, where $\mu = +$, $-$, or $(0)$ have to be calculated numerically.
However, in the special case $P_d = 0$ where the domain walls act as decoupled 1D wires,
these functions have a well-known analytical form~\cite{Voit1995, Schulz2000, Giamarchi2003, Gogolin2004}
\begin{align}
P^{\nu}_{a p, a'p'}(\mathbf{q}, \omega) &=  \delta_{a a'} \delta_{p p'} \frac{N}{2 \pi \hbar v}
	\frac{\nu v q_{a p}}{\omega - \nu v q_{a p}}\,,
\label{eqn:P_nu_free}\\
P^{(0)}_{a p, a'p'}(\mathbf{q}, \omega) &=  \delta_{a a'} \delta_{p p'} \frac{N}{\pi \hbar v}
\frac{v^2 q^2_{a p}}{\omega^2 - v^2 q^2_{a p}}\,.
\label{eqn:P_free}
\end{align}

Let us make a few more definitions we will need later on.
The RPA dielectric function is the matrix with elements
\begin{equation}
\begin{split}
\epsilon_{a p, a' p'}(\mathbf{q}, \omega) =& \delta_{a a'} \delta_{p p'}
\\
\mbox{} &- \sum_{a'' p''} \tilde U_{a p, a'' p''}(\mathbf{q}) P^{(0)}_{a'' p'', a' p'}(\mathbf{q}, \omega) .
\label{eqn:epsilon_RPA}
\end{split}
\end{equation}
The matrix inverse of this dielectric function has elements
\begin{equation}
	\begin{split}
		\epsilon^{-1}_{a p, a' p'}(\mathbf{q}, \omega) =& \delta_{a a'} \delta_{p p'}
		\\
		\mbox{} &+ \sum_{a'' p''} \tilde U_{a p, a'' p''}(\mathbf{q}) P_{a'' p'', a' p'}(\mathbf{q}, \omega) \,,
		\label{eqn:epsilon_RPA_inverse}
	\end{split}
\end{equation}
where
\begin{equation}
	P_{a p, a' p'}(\mathbf q,\omega) = \sum_{a'' p''} P^{(0)}_{a p, a'' p''}(\mathbf q,\omega)
	\epsilon^{-1}_{a'' p'', a' p'}(\mathbf{q}, \omega)
\label{eqn:P}
\end{equation}
is the total polarization, i.e.,
the density response function with respect to the external potential:
\begin{equation}
	n_{a p} = \sum_{a', p'} P_{a p, a' p'}(\mathbf q, \omega)
	e\Phi_{\mathrm{ext}, a' p'}(\mathbf{q}, \omega)\,.
\end{equation}
The various polarization matrices may be easily converted to the conventional 2D basis of reciprocal lattice vectors as follows:
\begin{equation}
	P^{\mu}_{\mathbf{G} \mathbf{G}'}(\mathbf{q}, \omega) = \frac{1}{L_{\bot}}\, 
	\sum_{a a'} \sum_{p p'} X_{\mathbf{G}, a p} X_{\mathbf{G}', a' p'}
	P^{\mu}_{a p, a' p'}(\mathbf{q}, \omega) \,,
\label{eqn:polarization_2D}
\end{equation}
with factors $X_{\mathbf{G}, a p}$ defined by Eq.~\eqref{eqn:X}.
In turn, the 2D dielectric function and its inverse are given by
\begin{align}
	\epsilon_{\mathbf{G} \mathbf{G}'}(\mathbf{q}, \omega) &= \delta_{\mathbf{G} \mathbf{G}'}
	- \tilde{U}(\mathbf{q} + \mathbf{G}) P^{(0)}_{\mathbf{G} \mathbf{G}'}(\mathbf{q}, \omega) \,,
	\label{eqn:epsilon_2D}\\
	\epsilon^{-1}_{\mathbf{G} \mathbf{G}'}(\mathbf{q}, \omega) &= \delta_{\mathbf{G} \mathbf{G}'}
	+ \tilde{U}(\mathbf{q} + \mathbf{G}) P_{\mathbf{G} \mathbf{G}'}(\mathbf{q}, \omega) \,.
	\label{eqn:epsilon_inverse_2D}
\end{align}

The plasmon dispersion can be found by locating the poles of the inverse dielectric function.
This is easy to do in the low-frequency, long-wavelength regime where it is sufficient to keep only the $p = p' = 0$ terms
so that $\epsilon^{-1}_{a p, a' p'}$ can be organized into a $3 \times 3$ matrix.
In the limit of vanishing deflection and short-range interaction,
where the links are decoupled, this matrix is diagonal:
\begin{equation}
	\epsilon^{-1}_{a 0, a' 0}(\mathbf{q}, \omega) = \delta_{a a'}
	\frac{\omega^2 - v^2 q^2_{a}}{\omega^2 - v_p^2 q^2_{a}}\,,
	\label{eqn:epsilon_decoupled}
\end{equation}
where
\begin{equation}
	v_p = \frac{v}{K} 
	\label{eqn:v_p}
\end{equation}
is the 1D plasmon velocity (derived in Sec.~\ref{sub:bosonization} below) and
\begin{equation}
	K(q) = 
	\left[1 + \frac{N}{\pi \hbar v}\,  \tilde{U}(q)\right]^{-1/2}
	\label{eqn:K}
\end{equation}
is the dimensionless function mentioned in Sec.~\ref{sec:Introduction}.
Note that for a short-range interaction, $v_p$ and $K$ are constants independent of $q$.
Accordingly, the long-wavelength plasmon dispersion consists of three acoustic branches:
\begin{equation} 
	\omega(\mathbf{q}) = v_a({\mathbf{q}}) q
	\label{eqn:2d_plasmon_short-range}
\end{equation}
with velocities $v_a = (q_a / q) v_p$ that depend on the direction of vector $\mathbf{q}$.
These velocities are modified if $P_d$ or the interaction range are nonzero
but as long as the interaction is screened, the three acoustic branches remain.

In the strongly interacting limit $K\ll 1$, the fastest acoustic branch
dominates the spectral weight and has velocity
\begin{equation} 
	v_s = \sqrt{\frac{D}{2D_\mathrm{max}}}v_p,
	\label{eqn:v_s}
\end{equation}
where
\begin{equation}
	D = 2 \pi e^2	N \sum_l \int \frac{d^2\! k}{(2\pi)^2}\, 
	{v}_{l x}^2(\mathbf k) \delta\left[\varepsilon_l(\mathbf k) - \varepsilon_F\right]
	\label{eqn:drude}
\end{equation}
is the Drude weight,
with $\mathbf{v}_l(\mathbf{k}) = \hbar^{-1} \nabla_{\mathbf{k}} \varepsilon_l$ the single-particle velocity of band $l$.
The Drude weight has its maximum value
\begin{equation}
D_\mathrm{max} = \sqrt{3}\, \kappa \alpha N\, \frac{v^2}{L}
\label{eqn:D_max}
\end{equation}
at full forward scattering $P_f = 1$ and decreases with decreasing $P_f$.
Here
\begin{equation}
	\alpha = \frac{e^2}{\kappa\hbar v}
\end{equation}
is the dimensionless parameter that characterizes the strength of the Coulomb interaction.
If the interaction is unscreened,
then the plasmon excitations with the largest spectral weight have a $\sqrt{q}\,$ dispersion
typical of 2D conductors:
\begin{equation} 
	\omega(q) \simeq \sqrt{\frac{2}{\kappa}\, D q}\,, \qquad q L \ll 1.
\label{eqn:2d_plasmon}
\end{equation}
The remaining two low-frequency dispersion branches are acoustic.
They have very small spectral weight, see below.

If $P_f < 1$, the Drude weight varies periodically with $\varepsilon_F$,
as seen in Fig.~\ref{fig:drude_weight}(a,right).
For example, if $v = 10^8\,\mathrm{cm} / \mathrm{s}$ and $L = 200\,\mathrm{nm}$,
the oscillations with Fermi energy have a period of $\Delta \varepsilon_F = 2\pi\hbar v/(3L) = 7\,\mathrm{meV} = 1.7\,\mathrm{THz} \times h$.
In terms of the electron density per area,
the oscillation period is $\Delta n_{\mathrm{2D}} = 2 N / A_0 = 2.3 \times 10^{10}\,\mathrm{cm}^{-2}$.
Here
\begin{equation} 
	A_0 = L_{\bot} L = \frac{\sqrt{3}}{2}\, L^2
	\label{eqn:A_0}
\end{equation}
is the area of a unit cell of the network.
If detecting such rapid oscillations in experiment
proves challenging, then perhaps $\overline D$, the Drude weight averaged over one period, is more accessible. Our numerical results for $\overline D$ as a function of
$P_f$ are shown in Fig.~\ref{fig:drude_weight}(b),
together with the analytical approximation
\begin{equation}
\overline D \simeq \left(1 - \tfrac32 \sqrt{P_d}\, \right) D_\mathrm{max}\,,
\quad
P_d \ll 1\,,
\label{eqn:D_avg}
\end{equation}
derived in Appendix~\ref{app:sp}.
\begin{figure}[t]
	\begin{center}
		\includegraphics[width=2.75in]{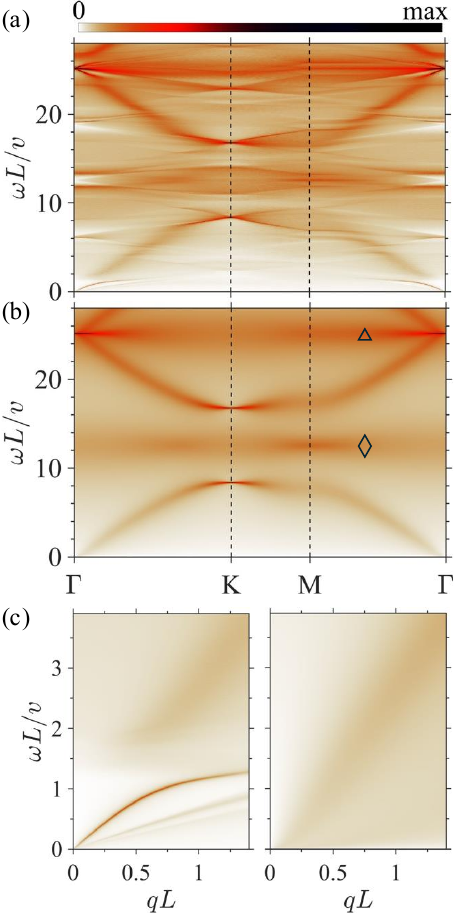}
	\end{center}
	\caption[RPA and PNM spectra for short-range interactions]
	{RPA and PNM spectra [Eq.~{\eqref{eqn:spectral_function}}] along high-symmetry lines for a short-range interaction. 
		(a) RPA spectrum [from Eqs.~{\eqref{eqn:polarization_as_sum}}--{\eqref{eqn:Psi2}}, {\eqref{eqn:epsilon_RPA}}] for $P_f = 0.4$ and $K = 0.25$.
		The Fermi energy $\varepsilon_F$ is indicated in Fig.~\ref{fig:drude_weight}(a).
		(b) Local PNM spectrum [from Eqs.~{\eqref{eqn:S}}, {\eqref{eqn:T}}--{\eqref{eqn:B}}] for the same parameters. The symbols label the frequencies $\omega_\triangle$ and $\omega_\Diamond$
		defined in Eqs.~\eqref{eqn:flat_band1} and \eqref{eqn:flat_band2}.
		(c) Enlarged views of the long-wavelength, low-frequency parts of the plots above. Note that at the $\mathrm{K}$ point $q L = 4\pi / 3$.
	}
	\label{fig:bands}
\end{figure}
\begin{figure}[t]
	\begin{center}
		\includegraphics[width=2.75in]{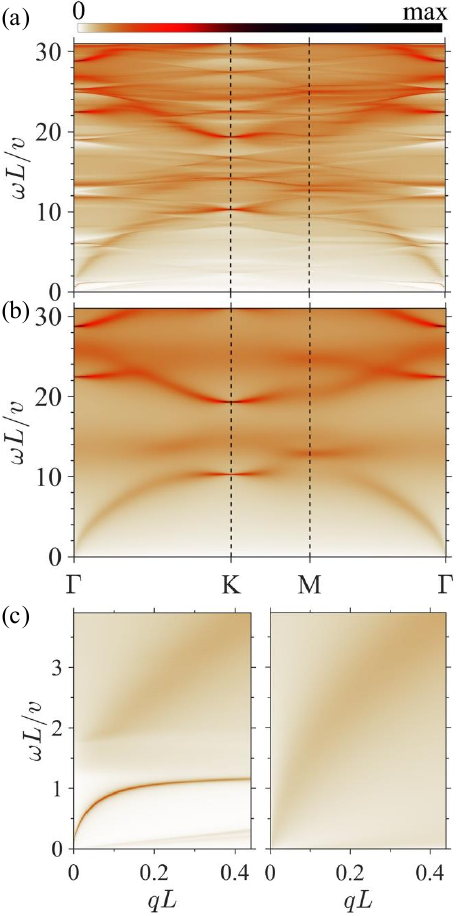}
	\end{center}
	\caption[RPA and PNM spectra for unscreened interactions]
	{RPA and PNM spectra along high-symmetry lines for an unscreened Coulomb interaction. Parameters: $\alpha N = 10$, $\ell = 0.01L$, $P_f$ and $\varepsilon_F$ are the same as in Fig.~\ref{fig:bands}.
		(a) RPA spectrum [from Eqs.~{\eqref{eqn:polarization_as_sum}}--{\eqref{eqn:Psi2}}, {\eqref{eqn:epsilon_RPA}}].
		(b) Non-local PNM spectrum [Eqs.~{\eqref{eqn:polarization_as_sum}}, {\eqref{eqn:epsilon_RPA}}, {\eqref{eqn:polarization3}}--{\eqref{eqn:lambda_s_a}}].
		(c) Enlarged views of the long-wavelength, low-frequency parts of the plots above.
	}
	\label{fig:bands2}
\end{figure}

Plasmon dispersion away from the asymptotic low-frequency limit has to be calculated numerically.
In general, this dispersion has both real and imaginary parts.
Its projection onto the space of real frequencies and momenta can be visualized by plotting the
trace of the spectral function:
\begin{equation} 
	\tr\, \mathbf A(\mathbf q,\omega) = -\im \sum_{ap} \epsilon^{-1}_{ap,ap}(\mathbf q,\omega) = -\im \sum_\mathbf{G} \epsilon^{-1}_{\mathbf{G} \mathbf{G}}(\mathbf q,\omega)
	\label{eqn:spectral_function}
\end{equation}
over a desired range of $\mathbf{q}$ and $\omega$ (assuming $\mathbf{q}$ stays within the moir\'e Brillouin zone).
We compute $\tr\, \mathbf A(\mathbf q,\omega)$ numerically 
using Eqs.~{\eqref{eqn:polarization_as_sum}}--{\eqref{eqn:Psi2}}, {\eqref{eqn:epsilon_RPA}}, and {\eqref{eqn:spectral_function}}.

The case of a short-range interaction [Eq.~\eqref{eqn:short_range_kernel}]
is illustrated by Fig.~\ref{fig:bands}(a).
The spectrum is quasi-periodic in frequency so we plot a range of $\omega$ only slightly larger than one period $2\pi v_p/L$. At low frequencies it consists of three linearly dispersing gapless modes, as predicted by Eq.~\eqref{eqn:2d_plasmon_short-range}.
They can be seen more clearly in an enlarged view of this region, Fig.~\ref{fig:bands}(c, left).
At higher frequencies, the plasmon modes typically lie in one of the numerous particle-hole continua and have a finite lifetime.
However, at high-symmetry points $\Gamma$, $\mathrm{K}$, and $\mathrm{K}'$,
there are dissipationless modes. Their presence can be explained by the existence of
two conserved quantities in the system, the total charge and the valley polarization,
see Sec.~\ref{sec:PNM}.
Also notable are weakly dispersive bands near frequencies
\begin{align}
	\omega_\triangle(\mathbf q) &= \frac{2 \pi v_p}{L}\,,
	\label{eqn:flat_band1} \\
	\omega_\Diamond(\mathbf q) &= \frac{\pi v_p}{L}\,.
	\label{eqn:flat_band2}
\end{align}
Similar bands are also predicted by the PNM, as labeled in Fig.~\ref{fig:bands}(b)
and discussed further in Sec.~\ref{sec:PNM}.

The RPA plasmon dispersion calculated for the unscreened Coulomb interaction
[Fig.~\ref{fig:bands2}(a)] is overall similar to that for the screened one [Fig.~\ref{fig:bands}(a)]
except the dominant low-frequency mode has the $\sqrt{q}\,$ rather than linear in $q$ dispersion,
cf. Figs.~\ref{fig:bands2}(c, left) and \ref{fig:bands}(c, left), in agreement with Eq.~\eqref{eqn:2d_plasmon}.
In addition, the four-fold band degeneracy near $\omega = \omega_\triangle$ found at the $\Gamma$-point is partially lifted in the unscreened case
for the reasons to be explained in Sec.~\ref{sec:PNM}.

While the RPA is a standard theoretical approach to study electron systems,
it is not completely reliable.
This approximation neglects vertex corrections to the polarization functions $P^{(0)}_{\mathbf{G} \mathbf{G}'}(\mathbf{q}, \omega)$,
which can produce exciton bound states outside the particle-hole continua and excitonic enhancements inside the continua.
For example, the latter can modify the power-law exponents $\pm 1 /2$ of the 1D van Hove singularities found near the edges of such continua.
Despite these shortcomings, the RPA should be a reasonable approximation
for mTBG  in the coherent regime because the vertex corrections are suppressed by the parameter $1 / N$, as alluded to in Sec.~\ref{sec:Introduction2}.
On the other hand, the PNM [Figs.~\ref{fig:bands}(b) and \ref{fig:bands2}(b)], on which we will focus next,
could be a good approximation in the incoherent regime.

%%%%%%%%%%%%%%%%%%%%%%%%%%%%%%%%%%%%%%%%%%%%%%%%%%%%%%%%%%%%%%%%
\section{Plasmon scattering at a junction} \label{sec:scattering}

A key ingredient of the PNM is the solution of the plasmon scattering problem at a single junction.
The problem is set up by treating the links attached to the node as semi-infinite 1D leads.
We start with briefly summarizing the standard bosonization approach and key equations for collective modes of such 1D systems.
Thereafter we show how to generalize these equations for the case of a six-lead junction and discuss the
solution of the scattering problem in hand.

\subsection{Plasmons of a wire} \label{sub:bosonization}

Consider an isolated wire containing a certain number $N$ of 1D propagating modes ($N = 4$ in mTBG). 
It is customary~\cite{Solyom1979, Voit1995, Schulz2000, Giamarchi2003, Gogolin2004} to describe the long-wavelength fluctuations of the electron density $n_k$ and the current $j_k$ of mode $k$ in terms of
densities of the right ($+$) and left ($-$) going fermions:
\begin{equation}
	n_{k} = n_{k}^{+} + n_{k}^{-}\,,
	\qquad
	j_{k} = e v \left(n_{k}^{+} - n_{k}^{-}\right)\,,
	\label{eqn:n_and_j_pm}
\end{equation}
which, in turn, can be expressed in terms of bosonic phase fields $\phi_{k}^{\pm}$:
\begin{equation}
	n_{k}^{\pm} = -\frac{1}{\pi} \left[\phi_{k}^{\pm}\right]'(x)\,.
	\label{eqn:phi_pm}
\end{equation}
We define nonchiral fields $\phi_k = \phi_{k}^{+} + \phi_{k}^{-}$ and 
$\Pi_k$ such that
\begin{equation}
	n_k = -\frac{1}{\pi} \phi_k'(x)\,,
	\qquad
	j_k = e v \Pi_k\,.
	\label{eqn:n_k_and_j_k}
\end{equation}
These fields satisfy the canonical commutation relations $[\phi_j(x), \Pi_k(x')] = i\delta_{j k} \delta(x - x')$.
The effective Hamiltonian of the wire is $H = H_{0} + H_{\mathrm{int}} + H_{\mathrm{ext}}\,$,
where
\begin{equation}
	H_{0} = \sum_{k = 1}^N
	\frac{\hbar v}{2\pi} \int dx
	\left[\pi^2\Pi_{k}^2(x) + (\phi_{k}')^2\right]
\end{equation}
is the kinetic energy,
\begin{equation}
	H_{\mathrm{int}} = \frac{1}{2} \sum_{j = 1}^N \sum_{k = 1}^N
	\int d x \int d y\, n_j(x) U(x - y) n_k(y)
\end{equation}
is the interaction energy, and
\begin{equation}
	H_{\mathrm{ext}} = \sum_{j = 1}^N
	\int d x\, e V_{\mathrm{ext}}(x) n_j(x)
\end{equation}
is the energy of interaction with an external potential $V_{\mathrm{ext}}(x)$.
Let us perform a linear transformation $\varphi_j(x) = \sum_k O_{j k} \phi_k(x)$
where $O_{j k}$ are constants forming an orthogonal matrix $\mathbf{O}$.
Such a transformation preserves the commutation relations among the fields.
The corresponding canonical momenta are denoted by $\varPi_k$.
If we choose all the elements of the first row of $\mathbf{O}$ to be equal, $O_{1 k} = 1 / \sqrt{N}$,
then the total density and current in the wire are expressed solely in terms of $\varphi_1$ and $\varPi_1$ fields:
\begin{equation}
	n(x) = -\frac{\sqrt{N}}{\pi}\, \varphi_1'(x)\,,
	\quad
	j(x) = \sqrt{N}\, e v \varPi_1(x)\,.
	\label{eqn:n_and_j}
\end{equation}
Therefore, $\varphi_1$ represents the charged excitation (plasmon) whereas the remaining $\varphi_k$'s with $k \neq 1$ describe neutral collective modes.
The Hamiltonian becomes
\begin{equation}
	\begin{split}
		H =& \frac{\hbar v}{2\pi} \sum_{k = 1}^N \int d x
		\left[\pi^2\varPi_k^2(x) + (\varphi_k')^2\right]
		\\
		&+ \frac{1}{2\pi^2} N \int d x \int d y\, \varphi_1'(x) U(x - y) \varphi_1'(y)
		\\
		&- \frac{1}{\pi} \sqrt{N} \int d x\, e V_{\mathrm{ext}}(x) \varphi_1'(x)\,.
	\end{split}
\end{equation}
The corresponding equations of motion are
\begin{align}
	\partial_t \varphi_k =& \pi v \varPi_k,
	\label{eqn:dot_varphi}\\
	\pi \partial_t \varPi_k =& v \varphi_k''(x)
	\nonumber \\
	&+ \delta_{1 k} \frac{1}{\pi \hbar} N
	\frac{\partial}{\partial x} \int  d y\, U(x - y) \varphi_1'(y)
	\nonumber \\
	&- \delta_{1 k} \frac{1}{\hbar} \sqrt{N}\,
	e V_{\mathrm{ext}}'(x)\,.
	\label{eqn:dot_Pi}
\end{align}
For a harmonic time dependence $e^{-i\omega t}$ of the fields, the equations for $k = 1$ entail
\begin{gather}
	-i\omega e n(x) + j'(x) = 0,
	\label{eqn:continuity}\\
	-i\omega j(x) = \frac{N e^2}{\pi \hbar} v F(x)\,,
	\label{eqn:Drude}\\
	F(x) = -V'(x)\,,
	\label{eqn:F}\\
	e V(x) = \frac{\pi\hbar v}{N}\, n(x) + \int U(x - y) n(y) d y 
	+ e V_\mathrm{ext}(x)\,.
	\label{eqn:Phi}
\end{gather}
The first term on the right-hand side of Eq.~\eqref{eqn:Phi} is the local chemical potential and the remaining two terms are the electrostatic energy per electron.
Therefore, $e V$ is the electrochemical potential, $V$ is the voltage,
and $e F$ is the electrochemical gradient. 
For simplicity, we will refer to $F$ as the electric field. 
The true electric field $E$ is related to $F$ via
\begin{equation}
	E(x) = F(x) + \frac{\pi \hbar v}{e N}\, n'(x) = F(x) + \frac{v^2}{\omega^2} F''(x)\,.
	\label{eqn:E}
\end{equation}
Combining Eqs.~\eqref{eqn:continuity}--\eqref{eqn:Phi} together, we obtain
\begin{equation} 
	\begin{split}
		F(x) &+ \frac{v^2}{\omega^2}
		\left[F''(x) + \frac{N}{\pi \hbar v} \int U'(x - y) F'(y) d y\right]
		= F_\mathrm{ext}(x) \,,
		\label{eqn:F_eq}
	\end{split}
\end{equation}
where $F_\mathrm{ext}(x) = -V_\mathrm{ext}'(x)$ is the external electric field.
We can solve Eq.~\eqref{eqn:F_eq} by taking the Fourier transform of both sides:
\begin{equation} 
	\tilde{F}(q) = \frac{\omega^2}{\omega^2 - v_p^2 q^2}
	\tilde{F}_\mathrm{ext}(q)\,,
	\label{eqn:F_q}
\end{equation}
where $v_p = v_p(q)$ is defined by Eq.~\eqref{eqn:v_p}.
Note that the inverse dielectric function of the wire is
$\epsilon^{-1}_\mathrm{1D}(q,\omega) \equiv \tilde{F}_\mathrm{ext}(q) / \tilde{E}(q)$.
Using Eqs.~\eqref{eqn:E} and \eqref{eqn:F_q}, it may be written as
\begin{equation}
	\epsilon^{-1}_\mathrm{1D}(q,\omega)
	= 1 - \frac{1 - K^2(q)}{q^2 - \frac{\omega^2}{v^2}\,K^2(q)}\, q^2
	= \frac{\omega^2 - v^2 q^2}{\omega^2 - v_p^2(q) q^2} \,,
	\label{eqn:1D_dielectric}
\end{equation}
which has a pole at the 1D plasmon momentum $q_p$ that satisfies the equation
\begin{equation}
\omega = v_p(q_p) q_p \,.
\label{eqn:q_p}
\end{equation}

\subsection{Plasmons of a junction} \label{sub:junction}

Consider now a junction of six leads labeled by $a = 1, 2, \ldots, 6$ in the clockwise direction,
see Fig.~\ref{fig:junction_circuit}.
(Note the departure from the earlier convention where $a$ had only three possible values.)
To model the dynamics of the plasmon modes we generalize Eqs.~\eqref{eqn:continuity}--\eqref{eqn:Phi} as follows:
\begin{gather}
	-i\omega e\, \mathbf{n}(r) + \mathbf{j}'(r) = 0,
\label{eqn:continuity_junction}\\
-i\omega\, \mathbf{j}(r) = \frac{N e^2}{\pi \hbar} v\, \mathbf{F}(r)\,,
\label{eqn:Drude_junction}\\
	\mathbf{F}(r) = \mathbf{F}_\mathrm{ext}(r)
	 - \frac{\pi\hbar v}{e N}\, \mathbf{n}'(r) 
- \frac{1}{e} \frac{d}{d r} \int\limits_{0}^{\infty} d s\, \mathbf{U}(r, s) \mathbf{n}(s) ,
\label{eqn:F_junction}
\end{gather}
where $r$ is the distance from the node.
The bold symbols represent six-component column vectors or $6 \times 6$ matrices.
For example, $\mathbf{U}(r, s)$ is the matrix of interaction kernels
\begin{equation}
	U_{a b}(r, s)
	= U\left(\sqrt{r^2 + s^2 - 2 r s \cos \beta_{a b}}\,\,\right),
	\label{eqn:dist}
\end{equation}
where $\beta_{a b}$ is the angle between leads $a$ and $b$.
We can eliminate $\mathbf{n}$ and $\mathbf{j}$ to obtain the equation for the field
\begin{equation}
	\mathbf{F}(r) + \frac{v^2}{\omega^2} \left[\mathbf{F}''(r) + \frac{N}{\pi \hbar v} \frac{d}{d r}
	\int\limits_0^\infty d s\, \mathbf{U}(r, s) \mathbf{F}'(s) \right] =
	\mathbf{F}_\mathrm{ext}(r)\,,
	\label{eqn:F_eq_multi}
\end{equation}
which holds at $r > 0$.
To establish the boundary conditions at $r = 0$,
we postulate that outgoing and incoming chiral currents
$\mathbf{j}^{\pm} = (e v \mathbf{n} \pm \mathbf{j}\,) / 2$ [Eq.~\eqref{eqn:n_and_j_pm}] 
are linearly related via a certain ``current-splitting'' matrix $\mathbf{M}$~\cite{Agarwal2010}:
\begin{equation}
	\mathbf{j}^+(0) = \mathbf{M}\, \mathbf{j}^-(0) .
	\label{eqn:bc2}
\end{equation}
Each column of $\mathbf{M}$ must sum to unity to ensure current conservation.
In the spirit of the RPA, we assume that
the matrix elements of $\mathbf{M}$ are scattering probabilities: $M_{a b} = |S_{0, a b}|^2$,
where $\mathbf{S}_0$ is the single-particle scattering matrix, Eq.~\eqref{eqn:S_0}.
This yields
\begin{equation}
	\mathbf{M} =
	\begin{pmatrix}
		R & P_d & 0 & P_f & 0 & P_d\\
		P_d & R & P_d & 0 & P_f & 0\\
		0 & P_d & R & P_d & 0 & P_f\\
		P_f & 0 & P_d & R & P_d & 0\\
		0 & P_f & 0 & P_d & R & P_d\\
		P_d & 0 & P_f & 0 & P_d & R
	\end{pmatrix}
	,
\end{equation}
where, in addition to the previously introduced parameters $P_f$ and $P_d$, we included a small backscattering probability $R > 0$ to regularize the matrix inversion later on.
Provided $R + P_f + 2 P_d = 1$, every column (and row) of $\mathbf{M}$ indeed sums to unity.

It is instructive to make a connection between our problem and
the physics of microwave circuits and transmission lines~\cite{Montgomery1987pom}.
To this end, we define the junction admittance matrix mentioned in Sec.~\ref{sec:Introduction}:
\begin{equation}
	\mathbf{Y} = \frac{2 N e^2}{h} \left(\,\mathbf{I}_6 + \mathbf{M}\,\right)^{-1}
	\left(\,\mathbf{I}_6 - \mathbf{M}\,\right) ,
	\label{eqn:Y_from_M}
\end{equation}
where $\mathbf{I}_6$ is the $6 \times 6$ identity matrix.
Matrix $\mathbf{Y}$ is chosen such that
\begin{gather}
	\mathbf{j}(0) = \mathbf{j}^+(0) - \mathbf{j}^-(0) = -\mathbf{Y}\: 
	\mathbf{V}_J,
	\label{eqn:bc}\\
	e\, \mathbf{V}_J = \frac{\pi \hbar v}{N}\, \mathbf{n}(0) \,.
	\label{eqn:n_jump_junction}
\end{gather}
These equations show that in the presence of a current, the electron densities on any pair of leads are in general different: $n_a(0) \neq n_b(0)$ if $a \neq b$.
The corresponding chemical potentials
are the elements of vector $e\, \mathbf{V}_J$.
Their differences $e {V}_{J, a} - e {V}_{J, b}$ give
the voltage drops between the leads.
Therefore, the admittance matrix $\mathbf{Y}$ relates the currents and voltages at the junction. 
This is why it
can be represented by a certain equivalent circuit (Fig.~\ref{fig:junction_circuit}).
In turn, the leads behave as transmission lines of admittance
\begin{equation}
	Y_* = \frac{2 N e^2}{h} K
	\label{eqn:Y_lead}
\end{equation}
each, see below.

Using Eq.~\eqref{eqn:Y_from_M}, one can check that every row and column of $\mathbf{Y}$ sums to zero,
which means $Y_0 = 0$ is an eigenvalue of $\mathbf{Y}$.
The full set of its eigenvalues $Y_m$, where $m$ is again the angular momentum modulo $6$,
is as follows:
\begin{subequations}
	\begin{align}
		Y_0 &= 0 \,,
		\label{eqn:Y_0}\\
		Y_{\pm 1} &= \frac{2 N e^2}{h}\, \frac{1 - R - P_d + P_f}{1 + R + P_d - P_f} 
		\to \frac{2 N e^2}{h}\, \frac{2 - 3 P_d}{3 P_d}\,,
		\label{eqn:Y_1}\\
		Y_{\pm 2} &= \frac{2 N e^2}{h}\, \frac{1 - R + P_d - P_f}{1 + R - P_d + P_f}
		\to \frac{2 N e^2}{h}\, \frac{3 P_d}{2 - 3 P_d}\,,
		\label{eqn:Y_2}\\
		Y_3 &= \frac{2 N e^2}{h}\, \frac{1 - R}{R} \to +\infty\,.
		\label{eqn:Y_3}
	\end{align} 
	\label{eqn:Y_eigenvalues}% <-- this prevents indentation of the next paragraph!
\end{subequations}
The corresponding eigenvectors are given by Eq.~\eqref{eqn:z_m}.
One familiar example of these formulas is encountered if we set $P_d$ to zero and $P_f$ to $1 - R$.
This yields $Y_m = 0$ for even $m$ and $Y_m \propto (1 - R) / R$ for odd $m$, which are the admittance eigenvalues predicted by the Landauer theory (see, e.g., Ref.~\onlinecite{Kamenev2001}) applied to decoupled 1D wires.
To model mTBG we take another limit, $R \to 0^+$,
where the rightmost expressions in Eq.~\eqref{eqn:Y_eigenvalues} are obtained.

Using Eqs.~\eqref{eqn:continuity_junction}, \eqref{eqn:Drude_junction}, and \eqref{eqn:bc}, we obtain the desired boundary condition for the field:
\begin{equation}
	\mathbf{Y}\, \mathbf{F}'(0) + \frac{i \omega}{v}\, \frac{2 N e^2}{h} \mathbf{F}(0) = \mathbf{0}\,. 
	\label{eqn:F_bc_multi}
\end{equation}
At this point we can state the problem we wish to address.
It is to solve Eq.~\eqref{eqn:F_eq_multi} with
$\mathbf{F}_{\mathrm{ext}} \equiv \mathbf{0}$,
the boundary condition~\eqref{eqn:F_bc_multi} at $r = 0$ and another boundary condition
\begin{equation}
	\begin{split}
		F_a(r) &\simeq  e^{-i q_p r} - S_{a a} e^{i q_p r},
		\\
		F_b(r) &\simeq -S_{a b} e^{i q_p r} , \quad a \neq b
	\end{split}
	\label{eqn:F_bc_far}
\end{equation}
at $r \to \infty$.
[Alternatively, instead of Eq.~\eqref{eqn:F_bc_multi}, we can
set $\mathbf{F}(0) = \mathbf{0}$ and
$\mathbf{F}_{\mathrm{ext}}(r) \propto \mathbf Y\mathbf{V}_J \delta\left(r - 0^+\right)$ in Eq.~\eqref{eqn:F_eq_multi}.
Physically,
such $\mathbf{F}_{\mathrm{ext}}$ represents a localized friction force induced by a resistive tunneling barrier.]
The coefficients $S_{a b}$ define the plasmon scattering matrix $\mathbf{S}$,
which has the representation
\begin{equation}
	\mathbf{S} = \sum_m e^{2 i \delta_m} \mathbf{z}_m \mathbf{z}_m^\dagger.
	\label{eqn:S_from_delta}
\end{equation}
Unlike the single-particle scattering phase shifts $\delta_{0, m}$ (Sec.~\ref{sub:single-particle_mTBG}), the plasmonic ones $\delta_m$ are in general complex numbers.
\begin{figure}[t]
	\begin{center}
		\includegraphics[width=2.75in]{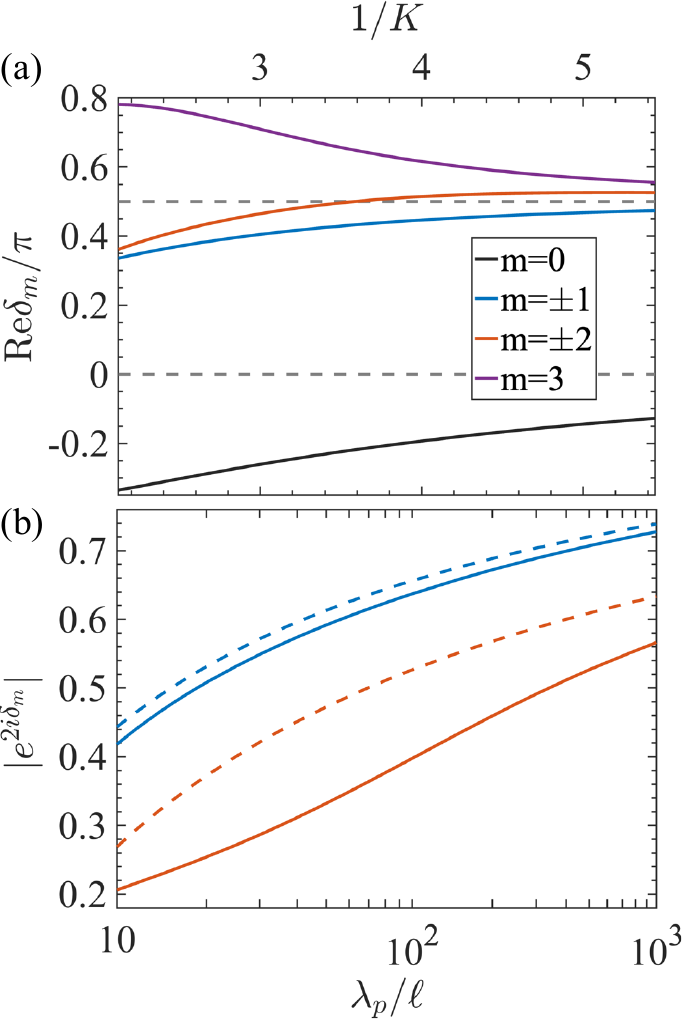}
	\end{center}
	\caption[Numerical results for scattering phase shifts.]
	{Scattering phase shifts for an unscreened Coulomb interaction versus the plasmon wavelength $\lambda_p$,
	obtained by numerically solving Eqs.~{\eqref{eqn:F_m}}--{\eqref{eqn:bc_m_far}}.
		Parameters are $P_f = 0.4$ and $\alpha N = 10$.
		(a) Real parts of the phase shifts in different angular momentum channels.
		The dashed lines show their asymptotic values.
		(b) Magnitude of the eigenvalues $e^{2 i \delta_m}$ of the $\mathbf{S}$ matrix for  $m = 1, 2$ (solid lines). The dashed lines represent Eq.~\eqref{eqn:S_analytical}.
		For $m = 0$ and $3$ channels, $\left|e^{2 i \delta_m}\right| = 1$ (not shown).
		\label{fig:six_channel}
	}
\end{figure}

In a channel with angular momentum $m$,
the fields have the form $\mathbf{F}(r) = F(r)\, \mathbf{z}_m$ so that
Eq.~\eqref{eqn:F_eq_multi} becomes
\begin{equation}
	F(r) + \frac{v^2}{\omega^2} \left[F''(r) + \frac{N}{\pi \hbar v} \frac{d}{d r}
	\int\limits_0^\infty d s\, U_m(r, s) F'(s) \right] =
	0\,,
\label{eqn:F_m}
\end{equation}
where the interaction kernel $U_m(r,s)$ is
\begin{equation}
	\begin{split}
		U_m(r,s) =& U(r - s)\\
		&+ 2\cos \frac{\pi m}{3}\, U\left(\sqrt{r^2 - r s + s^2}\, \right) \\
		&+ 2\cos \frac{2\pi m}{3}\, U\left(\sqrt{r^2 + r s + s^2}\, \right) \\
		&+ (-1)^m U(r + s).
	\end{split}
\label{eqn:U_m}
\end{equation}
In turn, the boundary conditions on $F(r)$ become
\begin{alignat}{2}
	&Y_m F'(r) +  \frac{i \omega}{v}\, \frac{2 N e^2}{h} F(r) = 0\,,& \quad
	&r = 0^+\,,
\label{eqn:bc_m}
\\
	&F(r) \simeq e^{-i q_p r} - e^{2 i \delta_m} e^{i q_p r},& &r \to \infty\,.
\label{eqn:bc_m_far}
\end{alignat}
Analytical solution of Eqs.~\eqref{eqn:F_m}--\eqref{eqn:bc_m_far}
for an arbitrary $U(r)$ is not possible.
An approximate solution can be obtained if we treat the interaction range $d_g$ as a small parameter.
In the leading order, we can replace $U_m(r, s)$ by $\tilde{U}(q_p) \delta(r - s)$,
which has the effect of changing the exact boundary condition~\eqref{eqn:bc_m} to
\begin{equation}
	Y_m F'(r) +  i q_p Y_* F(r) = 0\,,
	\qquad d_g \ll r \ll \lambda_p\,.
	\label{eqn:F_bc_m}
\end{equation}
Additionally, Eq.~\eqref{eqn:bc_m_far} is valid at all $r \gg d_g$.
Combining these equations and setting $d_g = 0$ to stay within the accuracy of the
approximation, we find
\begin{equation}
	e^{2 i \delta_m} = \frac{Y_* - {Y}_m}{Y_* + {Y}_m}.
\label{eqn:S_analytical}
\end{equation} 
Therefore, $m = 0$ and $m = 3$ channels are dissipationless, with $\delta_0 = 0$ and $\delta_3 = \pi/2 + i 0^+$.
In the $|m| = 1, 2$ channels where the $Y_m$'s [Eq.~\eqref{eqn:Y_eigenvalues}] are neither zero nor infinite,
phase shifts have imaginary parts $\im \delta_{m} > 0$, which indicate plasmon damping.

As long as Eq.~\eqref{eqn:S_analytical} applies,
$\mathbf{S}$ can be written in the form common in the circuit theory~\cite{Montgomery1987pom}:
\begin{equation}
	\mathbf{S} = \left(\,Y_*\mathbf{I}_6 + \mathbf{Y}\,\right)^{-1}
	 \left(\,Y_*\mathbf{I}_6 - \mathbf{Y}\,\right).
	\label{eqn:S}
\end{equation} 
Thus, in the absence of interactions, the lead admittance is $Y_* = 2 N e^2 / h$ and $\mathbf{S} = \mathbf{M}$, cf.~Eq.~\eqref{eqn:Y_from_M}.
In the opposite limit of strong local interactions, $K \to 0$,
the lead admittance $Y_*$ becomes vanishingly small; hence,
$\delta_m \to \pi/2$ for all $m \neq 0$ (assuming $P_d > 0$) while $\delta_0 = 0$ still.
Matrix $\mathbf{S}$ becomes unitary in this limit.
Its elements approach the asymptotic form
\begin{equation} 
	S_{a b} \simeq \frac{1}{3} - \delta_{a b},
\quad
K \ll 1\,,	
\label{eqn:S_classical}
\end{equation} 
which is independent of $\mathbf{Y}$.

In Appendix~\ref{app:scattering} we show that for screened Coulomb interactions
the next-order correction to Eq.~\eqref{eqn:S_analytical}
is linear in the parameter $q_p d_g \ll 1$.
It is equivalent to adding certain imaginary parts to the admittance eigenvalues:
\begin{equation}
	Y_m \to Y_m - i\Delta Y_m
	\label{eqn:Y_m_with_C_m}
\end{equation} 
with $\Delta Y_m \propto \left(1 - K^2\right)q_pd_g$.
This correction originates from finite-range interactions between
different links. One can think of it as an interlink capacitance in the equivalent circuit
of Fig.~\ref{fig:junction_circuit}.
In turn, for unscreened interactions, we find $\Delta Y_m \sim 1 / \mathcal{L}$,
where $\mathcal L = \ln(A / q_p \ell)$ assuming $q_p \ell \ll 1$.
Hence, the scattering amplitudes also approach those for short-range interactions
as $q_p$ decreases but this happens much slower, logarithmically.
This behavior is confirmed in Fig.~\ref{fig:six_channel}, where we plot our numerical solutions of Eqs.~{\eqref{eqn:F_m}}--{\eqref{eqn:bc_m_far}}
for $\delta_m$ as functions of the plasmon wavelength $\lambda_p = 2\pi / q_p$.
In Fig.~\ref{fig:six_channel}(a), we plot the real parts of the phase shifts for each angular momentum channel.
At long wavelength, $\re \delta_m$ approaches zero for $m = 0$
and $\pi / 2$ for $m \neq 0$, in accordance with Eq.~\eqref{eqn:S_analytical}.
Figure~\ref{fig:six_channel}(b) shows $\left|e^{2 i \delta_m}\right|$ for $m = 1$ and $2$ compared to the prediction of Eq.~\eqref{eqn:S_analytical}.
Note that $\left|e^{2 i \delta_m}\right| < 1$ for such $m$, indicating energy dissipation in these channels.

\section{Plasmonic network model}  \label{sec:PNM}

In this section, we first introduce the local PNM, where interactions between distant links are neglected
and the plasmon modes are determined from the scattering matrix at each node.
Using this framework, we compute the plasmon spectrum for a short-range interaction.
We then discuss the non-local PNM, which is valid for arbitrary interaction kernels, 
and compare its predictions with those of the local approximation for the case of unscreened Coulomb interactions. 
Finally, we employ the non-local PNM to derive an effective medium theory that describes the system
in the low-frequency, long-wavelength regime.

\subsection{Local PNM}
\label{sub:local_PNM}

In the local PNM, the construction of the eigenproblem for plasmon modes follows the same steps as in
the single-particle case, Sec.~\ref{sub:single-particle_mTBG}.
The equation to solve is
\begin{equation}
\mathbf{T}\, {\mathbf{n}} = e^{-i q_p L}\, {\mathbf{n}} ,
\label{eqn:eigenproblem}
\end{equation}
where
\begin{equation}
{\mathbf n}
 = \left(n_{1-}, n_{2+}, n_{3-}, n_{1+}, n_{2-}, n_{3+}\right)^T
\end{equation}
is the six-component column vector of plasmon amplitudes on the three links of the unit cell and two propagation directions.
Keeping with Sec.~\ref{sec:scattering}, we also use a numerical index $a$ to label the components of ${\mathbf{n}}$ and other six-vectors, such that
``$1-$'' is $a = 1$, ``$2+$'' is $a = 2$, and so on until ``$3+$'' which is $a = 6$,
see also Fig.~\ref{fig:junction_circuit}.
Matrix $\mathbf{T}$ is given by
\begin{gather}
\mathbf{T} =
\begin{pmatrix}
	0 & \mathbf{I}_3 \\ \mathbf{I}_3 & 0
\end{pmatrix}
\bm{\Lambda}^{-1} \mathbf{S}\, \bm{\Lambda}, \label{eqn:T}
\\
\bm{\Lambda} = \operatorname{diag} \left(1,e^{-iq_2 L}, 1, e^{-i q_1 L}, 1, e^{-i q_3 L}\right) ,
\end{gather}
where $\mathbf{q}$ is the crystal momentum and $q_a = \mathbf q \cdot \hat{\mathbf{l}}_a$.
Once the eigenvalue $e^{-i q_p L}$ [Eq.~\eqref{eqn:eigenproblem}] and so $q_p$,
which is in general a complex number, are found,
the plasmon dispersion is obtained from Eq.~\eqref{eqn:q_p}.
Alternatively, we can visualize this dispersion
by plotting the trace of the spectral function $\tr \mathbf A(\mathbf q,\omega) = -\im  \tr \bm{\epsilon}^{-1}(\mathbf q, \omega)$ over a desired range of real $\mathbf{q}$ and $\omega$.
The inverse dielectric function $\bm{\epsilon}^{-1}$ has matrix elements
\begin{align}
\epsilon^{-1}_{a p, a' p'}(\mathbf{q}, \omega) &= \delta_{aa'}\delta_{pp'}
 + \tilde\phi_{\mathrm{ind}, a p, a' p'}
\,,
\label{eqn:epsilon_local}\\
\tilde\phi_{\mathrm{ind}, a p, a' p'} &= \frac{1}{L}\, \int\limits_0^L d x
 e^{-i q_{a p} x} \int\limits_0^L d x' e^{i q_{a' p'} x'}
 \phi_{a a'}(x, x') \,,
\label{eqn:P_local}
\end{align}
with function $\phi_{a a'}(x, x')$ given by the following formula derived in Appendix~\ref{app:local}:
\begin{align}
\phi_{a a'} &= \left(1 - K^2\right)\left(\phi_{a a'}^{(1)} + \phi_{a a'}^{(2)} + \phi_{a a'}^{(3)}\right)\,, \label{eqn:P_local_r}
\\
\phi_{a a'}^{(1)} &= - \delta_{a a'}
\left[\frac{i}{2} q_p e^{i q_p |x - x'|} + \delta(x - x') \right] ,
\notag\\
\phi_{a a'}^{(2)} &= -\frac{i}{2} q_p \left[
B_{a+, a'+} e^{i q_p (x - x')} + B_{a-, a'-} e^{-i q_p (x - x')}\right] ,
\notag\\
\phi_{a a'}^{(3)} &= -\frac{i}{2} q_p  \left[
B_{a+, a'-} e^{i q_p (x - x')} + B_{a-, a'+} e^{-i q_p (x - x')}\right] , \notag
\end{align}
where $K = K(q_p)$, with $q_p$ being the plasmon momentum at frequency $\omega$ [Eq.~\eqref{eqn:q_p}].
The term $\phi_{a a'}^{(1)}$ represents the plasmonic response of the links. It can be obtained by taking the Fourier transform of Eq.~\eqref{eqn:1D_dielectric}.
The terms $\phi_{a a'}^{(2)}$ and $\phi_{a a'}^{(3)}$ represent the effect of scattering at the nodes in the directions that are the same or opposite to the single-particle scattering channels, respectively, with the coefficients $B_{a b}$ forming the $6 \times 6$ matrix
\begin{equation}
\mathbf{B} = \left(e^{-i q_p L}\, \mathbf{I}_6 - \mathbf{T}\right)^{-1}
\mathbf{T} \,.
\label{eqn:B}
\end{equation}
As in Sec.~\ref{sec:RPA}, we add a small imaginary part $\gamma_d$ to $\omega$ to regularize the spectral function.

A representative plot of the spectral function is shown in Fig.~\ref{fig:bands}(b)
for the short-range interaction case [Eqs.~{\eqref{eqn:spectral_function}}, {\eqref{eqn:S}}, {\eqref{eqn:T}}--{\eqref{eqn:B}}].
This plot illustrates the regime of a fairly strong single-particle deflection, $P_f = 0.4$.
As in the RPA [Fig.~\ref{fig:bands}(a)], the plasmon spectrum is quasi-periodic in frequency, with six bands per period.
The low-frequency spectrum consists of two diffusive branches with dispersions 
\begin{align}
	\omega(\mathbf q) &= -i \mathcal{D}_p q^2 ,
	\label{eqn:omega_diffusive_1}\\
	\omega(\mathbf q) &= -i \mathcal{D}_0 q^2 ,
	\label{eqn:omega_diffusive_2}
\end{align} 
with the diffusion coefficients
\begin{equation}
	\mathcal{D}_p = \frac12 \frac{v_p^2}{\gamma}\,,
	\quad
	\mathcal{D}_0 = \frac12 \frac{v^2}{\gamma}\,,
	\quad
\label{eqn:D_p}
\end{equation}
and a doubly-degenerate relaxational branch
\begin{equation}
\omega(\mathbf q) = (\pi - 2 \delta_1) \frac{v_p}{L}
\approx -i \gamma,
\label{eqn:omega_relaxational}
\end{equation}
where
\begin{equation}
\gamma = 2\, \frac{Y_*}{Y_1}\, \frac{v_p}{L}
       = \frac{6 P_d}{2 - 3 P_d}\, \frac{v}{L}
\label{eqn:gamma}
\end{equation}
is the relaxation rate. Here we assume, for simplicity, that
this relaxation rate is relatively small, $\gamma \ll v_p / L$,
which requires $Y_* \ll Y_1$.
However, $\gamma = 0.41 v_p / L$ in Fig.~\ref{fig:bands}.
A magnified view of the low-frequency region is shown in Fig.~\ref{fig:bands}(c, right).
The periodic replicas of the diffusive branches produce the sharp modes at $\omega = \omega_\triangle \equiv 2 \pi v_p / L$ [Eq.~\eqref{eqn:flat_band1}] at the $\Gamma$-point.
At larger $q$, these four bands evolve into two linearly dispersing bands $\omega_\pm(\mathbf q)$ [Eq.~\eqref{eqn:omega_pm} below] plus a two-fold degenerate flat band.
The remaining two bands are also flat and degenerate.
Their frequency is
$\omega(\mathbf q) \simeq 2(\pi - \delta_2) v_p / L$,
so that $\re \omega(\mathbf q) = \omega_\Diamond$,
Eq.~\eqref{eqn:flat_band2}.
The two dispersing modes resharpen at point $\mathrm{K}$ (and $\mathrm{K}'$, not shown).
The existence of dissipationless modes at the $\Gamma$, $\mathrm{K}$, and $\mathrm{K}'$ points 
is due to lack of dissipation in the $m=0$ and $m=3$ channels,
which in turn follows from conservation of the total charge and valley polarization in mTBG, respectively.
Comparing Figs.~\ref{fig:bands}(a) and \ref{fig:bands}(b), we conclude that the PNM and the RPA results agree in terms of their gross features but the former predicts a greater damping of all of them.
\begin{figure}[t]
	\begin{center}
		\includegraphics[width=2.75in]{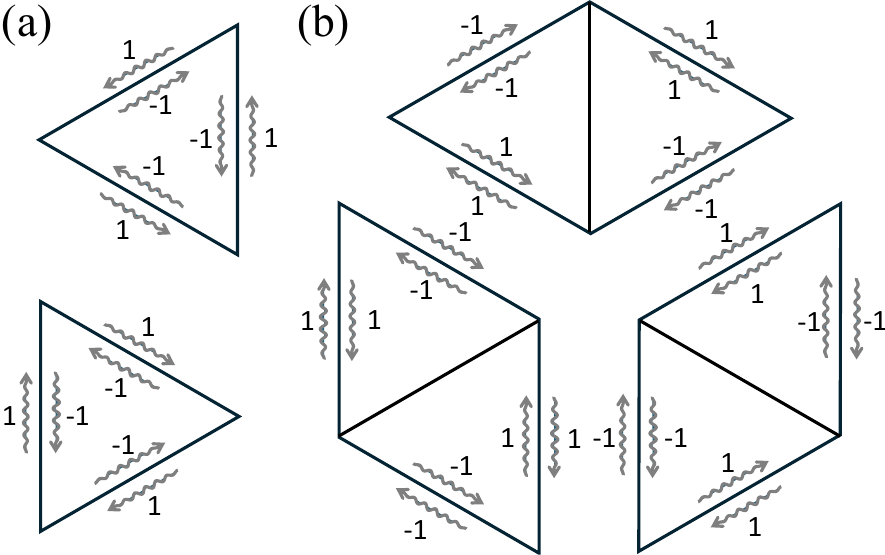}
	\end{center}
	\caption{
		Localized basis states corresponding to the flat bands of (a) $\omega_\triangle$, Eq.~\eqref{eqn:flat_band1} and (b) $\omega_\Diamond$, Eq.~\eqref{eqn:flat_band2}.
		The numbers next to the arrows denote amplitudes of the plasmon waves for such propagation directions.}
	\label{fig:localized_modes}
\end{figure}

Lastly, it is instructive to consider the strong-interaction limit $K \to 0$ with $\mathbf{S}$ given by Eq.~\eqref{eqn:S_classical},
for which Eq.~\eqref{eqn:eigenproblem} can be solved analytically.
Since $\mathbf{S}$ becomes unitary in this limit, all the modes are dissipationless.
Equations~\eqref{eqn:flat_band1} and \eqref{eqn:flat_band2}
turn out to be exact.
As stated above, each of these nondispersive modes is two-fold degenerate,
yielding the total of four bands.
To elucidate their spatial structure, we can construct localized (``Wannier'') basis functions from the Bloch functions.
For $\omega_\triangle$-modes, the most compact Wannier functions are confined to triangles, as illustrated in Fig.~\ref{fig:localized_modes}(a).
For $\omega_\Diamond$-modes, the Wannier functions are localized on rhombi, Fig.~\ref{fig:localized_modes}(b).
Although three such rhombi are shown in Fig.~\ref{fig:localized_modes}(b), there is a linear dependence between the corresponding Wannier functions.
The remaining two modes have a gapless dispersion (modulo $2\pi v_p / L$),
which is linear near the $\Gamma$ point:
\begin{equation}
	\omega_{\pm}(\mathbf q) = \pm \frac{v_p}{L}\,
	\arccos\left(\frac{1}{3} \sum_{a = 1}^3
	\cos q_a L \right)
	\simeq \pm \frac{v_p\, |\mathbf{q}|}{\sqrt{2}} ,
	\quad |\mathbf{q}| L \ll 1 .
	\label{eqn:omega_pm}
\end{equation} 
If $K$ is small but nonzero, there is a
crossover from Eqs.~\eqref{eqn:omega_diffusive_1} and \eqref{eqn:omega_relaxational} to Eq.~\eqref{eqn:omega_pm}, which occurs
when $v_p |\mathbf{q}|$ exceeds $\gamma$.
Furthermore, the analysis presented in Appendix~\ref{app:non-local} shows that
the range of validity of Eq.~\eqref{eqn:omega_pm} is restricted by the inequality $v |\mathbf{q}| \ll \gamma$.
At larger $|\mathbf{q}|$, the plasmon scattering at the junctions becomes a small perturbation.
The system behaves in the first approximation as a collection of nearly decoupled 1D wires
with three independent branches of acoustic plasmons, Eq.~\eqref{eqn:2d_plasmon_short-range}.

\subsection{Non-local PNM and effective medium theory}
\label{sub:nonlocal_PNM}

The local PNM described above is well suited for short-range interactions.
However, for unscreened Coulomb interactions, coupling between plasmons on distant links should also be considered.
In Appendix~\ref{app:non-local}, we derive general equations for the PNM that include such non-local couplings.
We find the formula for the dielectric function,
which is similar to that of the RPA [Eq.~\eqref{eqn:epsilon_RPA}] except
the irreducible polarization function $P^{\nu}_{ap,a'p'}$ of valley $\nu$ is now
\begin{equation}
	\begin{split}
		P^{\nu}_{a p, a'p'}(\mathbf q, \omega) =&  \delta_{a a'} \delta_{p p'} \frac{N}{2 \pi \hbar v}
		\frac{\nu v q_{a p}}{\omega - \nu v q_{a p}}
		\\
		\mbox{} &+ i \gamma \frac{N}{2 \pi \hbar v}
		\frac{\mathcal{G}^{\nu}_{a a'}(\mathbf{q}, \omega)}{\omega - \nu v q_{a p}} \,
		\frac{\nu v q_{a' p'}}{\omega - \nu v q_{a' p'}} \,.
	\end{split}	
\label{eqn:polarization3}
\end{equation}
This equation resembles Eq.~(25) of Ref.~\onlinecite{Kamenev2001} for an isolated 1D wire.
The first term in Eq.~\eqref{eqn:polarization3} is the result for uncoupled wires [Eq.~\eqref{eqn:P_nu_free}], the second term is due to scattering, $\mathcal{G}^{\nu}_{a a'}$ are the elements of a $3 \times 3$ matrix
\begin{equation}
	\begin{split}
		\mathbfcal{G}^{\nu}(\mathbf q, \omega) =& \frac{1}{3 + \sum_{a} \left(2\lambda^{\nu}_{a} + \lambda^{\nu}_{a} \lambda^{\nu}_{a + 1}\right)}\\
		\mbox{} & \times
		\begin{pmatrix}
			-2 - \lambda^{\nu}_{2} - \lambda^{\nu}_{3} & 1 + \lambda^{\nu}_{3} & 1 + \lambda^{\nu}_{2} \\
			1 + \lambda^{\nu}_{3} & -2 - \lambda^{\nu}_{3} - \lambda^{\nu}_{1} & 1 + \lambda^{\nu}_{1}\\
			1 + \lambda^{\nu}_{2} & 1 + \lambda^{\nu}_{1} & -2 - \lambda^{\nu}_{1} - \lambda^{\nu}_{2}
		\end{pmatrix} ,
	\end{split}
	\label{eqn:G_matrix}
\end{equation}
where
\begin{equation}
	\lambda^{\nu}_{a} = i \frac{\gamma L}{2 v}\,
	\cot \left(\frac{\omega - \nu v q_a}{2 v}\, L\right) \,.
	\label{eqn:lambda_s_a}
\end{equation}
As previously, $\omega \rightarrow \omega + i\gamma_d$.

\begin{figure}[t]
	\begin{center}
		\includegraphics[width=2.75in]{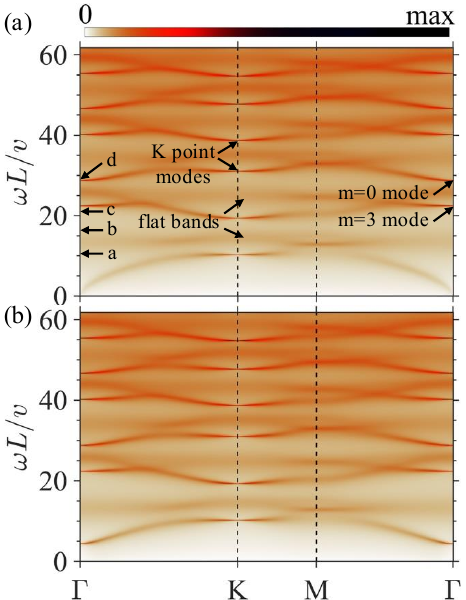}
	\end{center}
	\caption[Local vs. non-local PNM spectra]
	{PNM spectra [Eq.~{\eqref{eqn:spectral_function}}] along high-symmetry lines for unscreened Coulomb interactions for the same parameters as in Fig.~\ref{fig:bands2} but over a larger frequency range. 
	(a) Non-local PNM [Eqs.~{\eqref{eqn:polarization_as_sum}}, {\eqref{eqn:epsilon_RPA}}, {\eqref{eqn:polarization3}}--{\eqref{eqn:lambda_s_a}}].
	The labels indicate various features of the band structure as well as the frequencies a-d used to calculate the images in Fig.~\ref{fig:images}.
	 (b) Local PNM [Eqs.~{\eqref{eqn:T}}--{\eqref{eqn:B}}]. The corresponding scattering phase shifts are plotted in Fig.~\ref{fig:six_channel}.}
	\label{fig:bands3}
\end{figure}

An example of a plasmon spectrum computed in the non-local PNM for unscreened interactions
[Eqs.~{\eqref{eqn:polarization_as_sum}}, {\eqref{eqn:epsilon_RPA}}, {\eqref{eqn:spectral_function}}, {\eqref{eqn:polarization3}}--{\eqref{eqn:lambda_s_a}}]
is shown in Fig.~\ref{fig:bands2}(b)
[with a magnified view of the low-frequency region in Fig.~\ref{fig:bands2}(c,right)].
This spectrum has the same quasi-periodic structure
and gross features
as the corresponding RPA result in Fig.~\ref{fig:bands2}(a).
Figure~\ref{fig:bands3} helps to compare the spectra of the non-local and local PNM over a range of frequencies containing several periods of the plasmonic band structure.
The two spectra are in a good agreement at high frequency.
Evidently, in this regime the local PNM is sufficient, meaning that the plasmon spectra can be computed by solving Eq.~\eqref{eqn:eigenproblem}
with a wavelength-dependent junction scattering matrix (cf. Fig.~\ref{fig:six_channel}),
neglecting coupling between distant links.
This can be explained by the rapid spatial oscillations of the charge density on the links
which weakens this kind of coupling.

A closer inspection of Figs.~\ref{fig:bands}(b) and \ref{fig:bands2}(b),
which correspond to short-range and unscreened interactions, respectively,
reveals that in the latter case the degeneracy of the sharp modes at the $\Gamma$ point is lifted. The $m = 0$ mode is blue shifted and the $m = 3$ mode is red shifted, see
Fig.~\ref{fig:bands3}(a).
Within the local PNM, the frequencies of these modes are determined by the solutions of the equations
\begin{equation}
q_p L = -2 \delta_m(q_p) + 2\pi n, \quad m = 0 \text{ or } 3,
\end{equation}
where $n > 0$ is an integer.
Our perturbation theory for the phase shifts $\delta_m$ (see Appendix~\ref{app:scattering}) indicates
that the mode splitting in terms of $q_p$ should scale as $1 / \mathcal{L}$.

At low frequency, the local PNM fails, erroneously predicting a gapped dispersion,
Fig.~\ref{fig:bands3}(b), whereas the non-local PNM spectrum is gapless, Fig.~\ref{fig:bands3}(a).
We can show that the latter obeys the correct $\sqrt{q}\,$-law
using the effective medium theory (EMT).
This is based on the assumption that
it is permissible to approximate the dielectric function matrix by a scalar,
\begin{equation}
	\epsilon_{\mathrm{EMT}}(\mathbf{q}, \omega) \equiv
	\epsilon_{\mathbf{0}\mathbf{0}}
	 = 1 - \tilde{U}(\mathbf{q}) \frac{1}{L_{\bot}}\, 
	\sum_{a,a'}	P^{(0)}_{a\, 0, a'\, 0}(\mathbf{q}, \omega) \,,
	\label{eqn:epsilon_00}
\end{equation}
cf.~Eqs.~\eqref{eqn:polarization_2D} and \eqref{eqn:epsilon_2D}.
The discarded matrix elements $\epsilon_{\mathbf{G} \mathbf{G}'}$ give only small corrections
if the interaction is long-range so that
$\tilde{U}(\mathbf{q} + \mathbf{G}) \ll \tilde{U}(\mathbf{q})$.
The general expression for $\epsilon_{\mathrm{EMT}}$ is given in Appendix~\ref{app:non-local}.
Expanding this to the order $O\left(q^2\right)$,
we find
\begin{equation}
	\epsilon_{\mathrm{EMT}}(\mathbf{q}, \omega) \simeq 1
	- \frac{\tilde{U}(\mathbf{q})}{\pi e^2}\, 
	\frac{D_\mathrm{max} q^2}{\left(\omega + i \gamma\right) \omega}
\,,	\quad v q \ll \gamma\,.
\label{eqn:epsilon_EMT}
\end{equation}
This implies that the network behaves as a continuous 2D medium with
conductivity $\sigma_\mathrm{EMT}$ of the Drude form:
\begin{equation} 
	\sigma_\mathrm{EMT}(\omega) = \frac{1}{\pi}\, \frac{D_\mathrm{max}}{\gamma - i \omega} \,.
	\label{eqn:sigma_2d_Drude}
\end{equation}
It can be shown that the EMT is valid for a wider class of interaction types if instead of $\tilde{U}(\mathbf{q})$ we use the effective 2D kernel
\begin{equation}
	\tilde{U}_\mathrm{EMT}(\mathbf q) = \frac{L_{\bot}}{9}
	\sum_{a,a'} \tilde{U}_{a\, 0, a'\, 0}(\mathbf q)\,.
\label{eqn:U_eff}
\end{equation}
For example, $\tilde{U}_{\mathrm{EMT}}(\mathbf{q}) = \frac{1}{3}\, L_{\bot} \tilde{U}(0)$
for short-range interactions.

Solving $\epsilon_{\mathrm{EMT}}(\mathbf{q}, \omega) = 0$, which is a quadratic equation for $\omega$,
we find two collective mode frequencies
\begin{equation}
	\omega(\mathbf q) = -\frac{i}{2} \gamma \pm 
	\left[
	\frac{D_\mathrm{max}}{\pi e^2}\, \tilde{U}_{\mathrm{EMT}}(\mathbf{q}) q^2 - \frac14 \gamma^2 \right]^{1/2} . 
	\label{eqn:disp_cm}
\end{equation}
For short-range interactions,
this formula agrees with our results in Sec.~\ref{sub:local_PNM}
in the strongly interacting limit $K\ll1$.
At $q \ll \gamma / v_p$, it reproduces the first diffusive mode and the relaxational mode, Eqs.~\eqref{eqn:omega_diffusive_1} and 
\eqref{eqn:omega_relaxational}, respectively.
At larger $q$, it predicts
\begin{equation} 
	\omega(\mathbf q) \simeq \pm\frac{v_p}{\sqrt 2} q - \frac{i}{2} \gamma \,,
	\qquad \frac{\gamma}{v_p} \ll q \ll \frac{\gamma}{v}\,,
\end{equation}
in agreement with Eq.~\eqref{eqn:omega_pm}.
For unscreened interactions with
$\tilde{U}_{\mathrm{EMT}}(\mathbf{q}) = 2\pi e^2 / \kappa |\mathbf{q}|$,
the fast diffusive mode [Eq.~\eqref{eqn:omega_diffusive_1}]
becomes superdiffusive,
\begin{equation} 
	\omega(\mathbf q) = -i \frac{2 D_\mathrm{max}}{\kappa \gamma} q	\,,
\end{equation}
which is a well-known result.
As $q$ increases beyond $\gamma^2 \kappa / D_\mathrm{max}$, this mode evolves into a propagating plasmon mode with the expected $\sqrt{q}$ dispersion:
\begin{equation} 
	\omega(\mathbf q) = \sqrt{\frac{2 D_\mathrm{max}}{\kappa} q}
	 - \frac{i}{2} \gamma \,.
\end{equation}
This is the gapless branch seen in Fig.~\ref{fig:bands3}(a).

To derive Eq.~\eqref{eqn:omega_diffusive_2} for the slower diffusive mode,
the expansion of Eq.~\eqref{eqn:epsilon_00} has to be pushed to the order $O\left(q^4\right)$.
The corresponding pole singularity of the inverse dielectric function is given by
\begin{equation}
	\epsilon^{-1}_{\mathrm{EMT}}(\mathbf{q}, \omega) \simeq
	\frac{Z(\mathbf{q})}{\omega + i \mathcal{D}_0 q^2}\,.
\label{eqn:slow_diffusion_pole}
\end{equation}
The residue of the pole,
\begin{equation}
	Z(\mathbf{q}) = \frac{i \pi}{48 \gamma^3}\,
	\frac{e^2 v^6 q^4 \cos^2 3\theta}{D_\mathrm{max} \tilde{U}_{\mathrm{EMT}}(\mathbf{q})}\,,
\label{eqn:Z}	
\end{equation}
i.e., the optical strength of this mode, has a six-fold anisotropy
and a small magnitude $\propto q^4 / \tilde{U}_{\mathrm{EMT}}(\mathbf{q})$.
Here $\theta$ is the angle between the vectors $\mathbf{q}$ and $\mathbf{l}_3$.
As explained earlier, this mode arises due to the absence of single-particle intervalley scattering,
which results in a conserved valley polarization.
Unfortunately, it is hardly possible to resolve this practically neutral, overdamped mode in
either of Figs.~\ref{fig:bands}(c,right), \ref{fig:bands2}(c,right), or \ref{fig:bands3}(a).

\section{Near-field imaging simulations} 
\label{sec:imaging}

Plasmons can be observed in real-space using scattering-type scanning near-field optical microscopy (s-SNOM)~\cite{Chen2019} that utilizes a movable probe with a sharp polarizable tip positioned near the sample.
The probe acts as a nano-optical antenna that launches or detects propagating surface plasmons.
Alternatively, plasmons can be launched by a fixed scatterer (``impurity'') located on the sample and then detected by the probe.
The near-field signal measured by the s-SNOM may depend on
the response of the sample and the probe-sample coupling in a complicated way~\cite{Chen2019, Jiang2016a}.
Here we use two common simplifying approximations that (i) the probe can be modeled as a point dipole positioned a distance $z_{\mathrm{tip}}$ away from the sample and
(ii) the measured signal $\mathcal{S}(\mathbf{r})$ is proportional to the difference in $z$-component of the electric field at the tip with and without the sample.
In experiment, $z_{\mathrm{tip}}$ varies in time as the probe oscillates at a finite tapping frequency $\Omega$, and the signal is demodulated 
at some high harmonic of this frequency to suppress the far-field background~\cite{Chen2019}.
We ignore this complication and assume $z_{\mathrm{tip}}$ to be fixed.
With these approximations, we show below that $\mathcal{S}(\mathbf{r})$ is given by the convolution of the inverse dielectric function of the sample with appropriate form factors, 
which have a characteristic spread of $\sim z_{\mathrm{tip}}$ in real space.

Let $\mathbf{r}_0$ be the in-plane position of the launcher.
Within the quasi-static approximation,
it produces an external potential $\Phi_{\mathrm{ext}}(\mathbf{r}_1 - \mathbf{r}_0)$ at a point $\mathbf{r}_1$ on the sample surface.
For tip-launched plasmons, we take
$\Phi_{\mathrm{ext}}(\mathbf{r}) = \Phi_{d}(\mathbf{r}, z_{\mathrm{tip}})$,
where
$\Phi_{d} \propto (r^2 + z^2)^{-3/2}$
is the in-plane potential of a point dipole oriented along the $z$ direction.
The 2D Fourier transform of this model potential is
$\tilde{\Phi}_{\mathrm{ext}}(\mathbf{q}) = e^{-q z_{\mathrm{tip}}}$.
For simplicity, we use the same form of $\tilde{\Phi}_{\mathrm{ext}}$ for plasmons launched by an impurity:
$\tilde{\Phi}_{\mathrm{ext}}(\mathbf{q}) = e^{-q r_{\mathrm{imp}}}$,
where $r_{\mathrm{imp}}$ is a characteristic radius of the launcher.
The induced in-plane potential $\Phi$ is given by
evaluating the convolution of $\Phi_{\mathrm{ext}}$ with
the inverse dielectric function $\bm{\epsilon}^{-1}$,
then subtracting off $\Phi_{\mathrm{ext}}$.
The Fourier transform of $\Phi$ can be expressed as
\begin{equation}
	\tilde{\Phi}({\mathbf{q}+ \mathbf G}, z = 0)
	= \sum_{\mathbf G'}\left[\epsilon^{-1}_{\mathbf{G} \mathbf{G}'}(\mathbf{q},\omega)-\delta_{\mathbf{G} \mathbf{G}'}\right]
	\tilde{\Phi}_{\mathrm{ext}}(\mathbf{q} + \mathbf{G}')
	e^{-i(\mathbf{q} + \mathbf{G}') \cdot \mathbf{r}_0} .
	\label{eqn:phi_tot}
\end{equation}
Within the quasi-static approximation, $\Phi$
satisfies the three-dimensional Laplace equation, so at an arbitrary point $z$ above the sample,
we have
\begin{equation}
	\tilde{\Phi}(\mathbf{q} + \mathbf{G}, z)
	= 
	\tilde{\Phi}({\mathbf{q} + \mathbf G}, 0) e^{-|\mathbf{q} + \mathbf{G}| z} .
	\label{eqn:phi_tot_z}
\end{equation}
By assumption, the near-field signal $\mathcal{S}(\mathbf{r})$ is proportional to the $z$-component of the electric field at the tip:
\begin{equation}
	\begin{split}
		\mathcal{S}(\mathbf{r}) =& -\frac{\partial}{\partial z_{\mathrm{tip}}} \Phi(\mathbf{r}, z_{\mathrm{tip}}) \\
		\mbox{} =&
		\sum_{\mathbf{G},\mathbf{G}'} \, \int \frac{d^2 q}{(2\pi)^2}
		e^{i (\mathbf{q} + \mathbf{G}) \cdot \mathbf{r}}
		e^{-i(\mathbf{q} + \mathbf{G}') \cdot \mathbf{r}_0}
		\\
		&\times {F}_d(\mathbf{q} + \mathbf{G})\, \left[\epsilon^{-1}_{\mathbf{G} \mathbf{G}'}(\mathbf{q},\omega)-\delta_{\mathbf{G} \mathbf{G}'}\right]
		\tilde{\Phi}_{\mathrm{ext}}(\mathbf{q} + \mathbf{G}') ,
	\end{split}
	\label{eqn:S_near-field}
\end{equation}
where
\begin{equation}
{F}_d(\mathbf{q}) = -\frac{\partial}{\partial z_{\mathrm{tip}}} \tilde{\Phi}_{d}(\mathbf{q}, z_{\mathrm{tip}})
= q e^{-q z_{\mathrm{tip}}} .
\label{eqn:F_d}
\end{equation}
This is the formula we use for our simulations except we add a small imaginary part $\gamma_d$ to $\omega$, for regularization purposes, see below.

\begin{figure}[t]
	\begin{center}
		\includegraphics[width=2.75in]{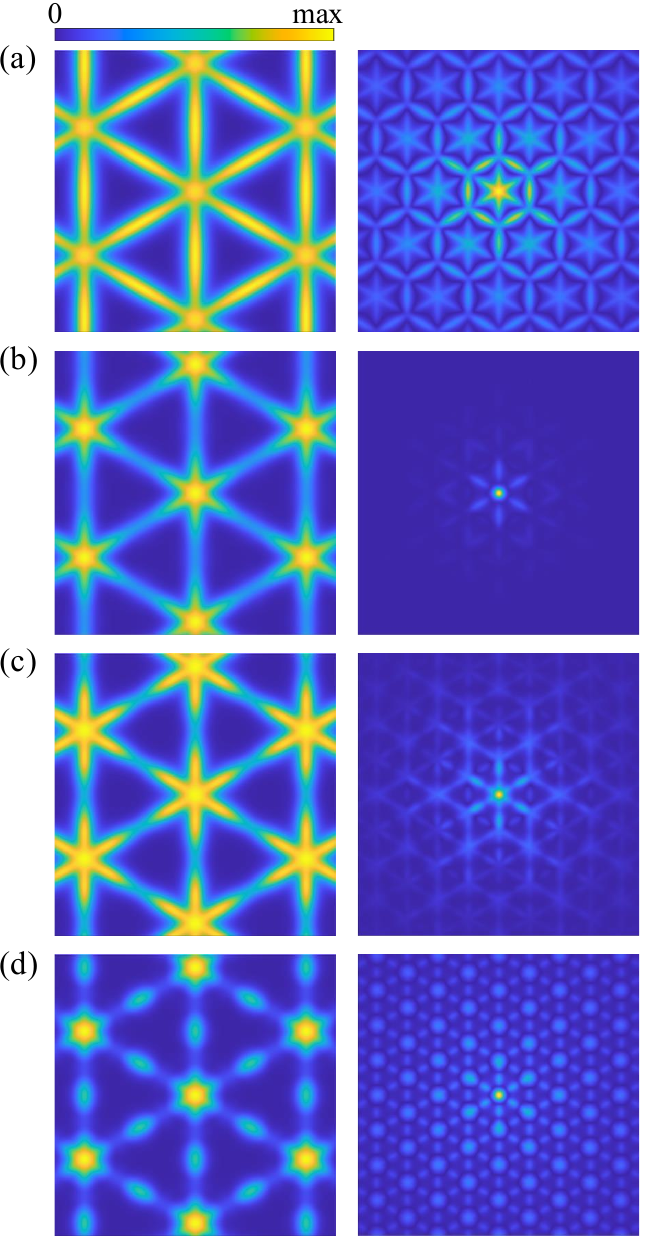}
	\end{center}
	\caption[Simulated near field images for PNM.]
	{Simulated near-field amplitude $|\mathcal{S}|$ within the PNM for plasmons launched by a tip (left) and impurity (right) 
	[Eqs.~{\eqref{eqn:polarization_as_sum}}, {\eqref{eqn:polarization_2D}}--{\eqref{eqn:epsilon_2D}}, {\eqref{eqn:polarization3}}--{\eqref{eqn:lambda_s_a}}, {\eqref{eqn:S_near-field}}--{\eqref{eqn:F_d}}].
	The corresponding frequencies for (a)-(d) are indicated in Fig.~\ref{fig:bands3}(a).
	Parameters: $z_{\mathrm{tip}} = 0.1 L$, $r_{\mathrm{imp}} = 0.1 L$.}
	\label{fig:images}
\end{figure}

Examples of calculations done with Eq.~\eqref{eqn:S_near-field} are shown in Fig.~\ref{fig:images}
where we use $\epsilon^{-1}_{\mathbf G \mathbf G'}(\mathbf q, \omega)$ computed within the PNM 
[Eqs.~{\eqref{eqn:polarization_as_sum}}, {\eqref{eqn:polarization_2D}}--{\eqref{eqn:epsilon_2D}}, {\eqref{eqn:polarization3}}--{\eqref{eqn:lambda_s_a}}]
with the same parameters as in Fig.~\ref{fig:bands3}(a),
and at the frequencies denoted there.
The left panels illustrate the scenario where the probe acts both as the launcher and the detector of the plasmons.
The near-field amplitude $|\mathcal{S}(\mathbf{r})|$ is periodic in the moir\'e superlattice, in agreement with Eq.~\eqref{eqn:S_near-field}, 
which contains only Fourier harmonics with momenta $\mathbf G - \mathbf G'$ if $\mathbf{r}_0 = \mathbf{r}$.
The signal can be understood as a result of interference between plasmons launched towards and reflected from the nearest nodes. Such interference should produce oscillations with
the period $\lambda_p / 2$,
which is roughly consistent with what we see in Fig.~\ref{fig:images}.
The links and nodes appear broadened with the characteristic width of the order of $z_{\mathrm{tip}} = 0.1 L$,
as we anticipated.

The right panels in Fig.~\ref{fig:images} are a simulation of the near-field signal produced by a local plasmon source in the sample plane.
In a uniform 2D conductor, such a source would produce a cylindrical wave,
$\mathcal{S}(\mathbf{r}) \sim H_0^{(1)}(q_p r)$ where 
$H_0^{(1)}(x)$ is the Hankel function of the first kind.
In our network system, we expect $q_p$ in this formula to be replaced,
in the first approximation,
by the magnitude $q$ of the crystal momentum $\mathbf{q} \parallel \hat{\mathbf{r}}$ at which the matrix element $\epsilon^{-1}_{\mathbf{G} \mathbf{G}'}(\mathbf{q}, \omega)$ in Eq.~\eqref{eqn:S_near-field} has a pole.
At a generic frequency $\omega$, the wavelength of such a wave will be direction-dependent.
Additionally, Eq.~\eqref{eqn:S_near-field} indicates that $\mathcal{S}(\mathbf{r})$ may acquire phase factors of the form $e^{i \mathbf{G} \cdot \mathbf{r}}$. Altogether, if $q$ is nonzero and distance $r$ is large such that $|q r| \gg 1$, we expect
$\mathcal{S}(\mathbf{r}) \sim f(\mathbf{r}) e^{i q r} / \sqrt{r}$,
where function $f(\mathbf{r})$ has the periodicity of a unit cell and some directional anisotropy.
If $q = 0$, then the formula in terms of the Hankel function gives infinity.
This is why the regularization parameter $\gamma_d$ is needed.
For small $\gamma_d$,
we expect $\mathcal{S}(\mathbf{r}) \sim H_0^{(1)}(q_d r)$ with $q_d \propto i\sqrt{\gamma_d}$
so that $|\mathcal{S}(\mathbf{r})| \sim \ln(l_d / r)$ at $r \ll l_d \propto 1 / \sqrt{\gamma_d}$.
A similar logarithmic decay
is expected if the frequency is in resonance with a sharp mode at the $\mathrm{K}$ point,
however, in this case it will be combined with oscillations of period $2 \pi / |\mathbf{K}| = 3 L / 2$.

\begin{figure}[t]
	\begin{center}
		\includegraphics[width=2.75in]{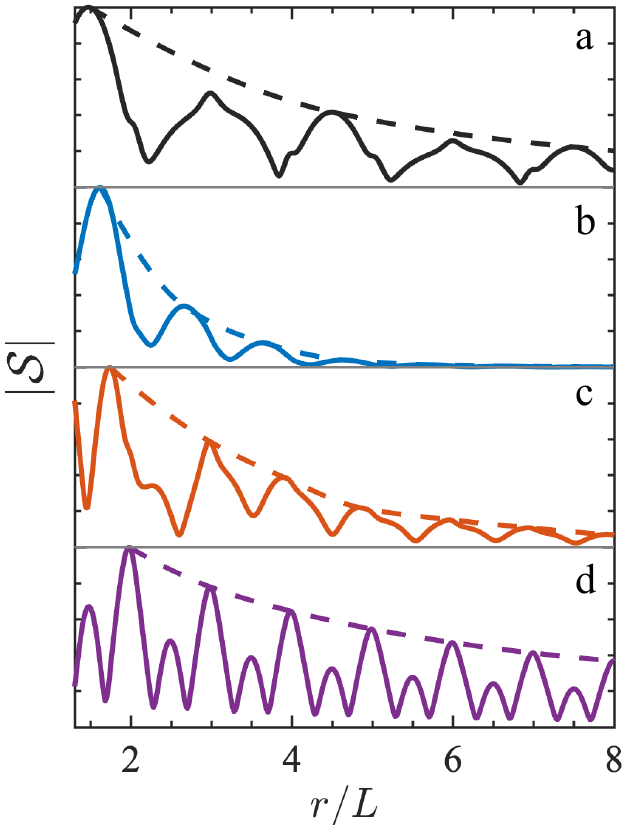}
	\end{center}
	\caption[Line profiles of simulated near field images for PNM.]
	{Line profiles along the central vertical cut through the images of Figs.~\ref{fig:images}(a-d, right) as functions of 
	the distance $r$ from the launcher. 
	 The dashed lines are guides for the eye.}
	\label{fig:line_profiles}
\end{figure}

To test these predictions, we examine the line profiles along the vertical, i.e., the lattice vector $\mathbf{l}_3$ direction, for Figs.~\ref{fig:images}(a-d, right).
Each profile exhibits a rapid initial decrease along the first couple links followed by a slower oscillatory decay to zero.
We select the latter region $r > 2L$ to analyze the functional form of the signal,
focusing on the envelope profiles indicated by the dashed lines in Fig.~\ref{fig:line_profiles}.
For the first example, plotted in Fig.~\ref{fig:line_profiles}(a), the frequency is in resonance with the sharp mode at the $\mathrm{K}$ point.
The slow decay of the envelope of this trace seems qualitatively consistent with the expected logarithmic law. The oscillations with period $3 L / 2$ are also evident.
In Fig.~\ref{fig:line_profiles}(b), the frequency lies in a plasmonic bandgap. The signal decays exponentially with $r$,
corresponding to a predominantly imaginary $q$.
In Fig.~\ref{fig:line_profiles}(c), the frequency is in resonance with a damped plasmonic mode.
This line profile exhibits oscillatory behavior along with a more gradual exponential decay,
consistent with a complex $q$.
Finally, in Fig.~\ref{fig:line_profiles}(d)
the frequency is in resonance with a sharp mode at the $\Gamma$ point.
Here the envelope of the trace appear to decrease logarithmically, as in Fig.~\ref{fig:line_profiles}(a);
however, the dominant oscillation period is the lattice constant $L$ rather than $3 L / 2$.

\section{Discussion} 
\label{sec:discussion}

In this work, we have formulated a theoretical model for plasmons
in the network of 1D states formed in a minimally twisted gapped bilayer graphene (mTBG).
Studying plasmon modes is crucially important for understanding low-energy properties of mTBG because electronic excitations of 1D systems are collective, e.g., the single-particle dispersion is unobservable.
Our calculations show that the plasmonic band structure of mTBG exhibits several unique features such as multiple gapless branches, flat bands, and dissipationless modes at high-symmetry points.
Since our goal was to elucidate the most essential attributes of the plasmonic response,
we worked within an oversimplified model that has only a handful of parameters, e.g.,
the bare Fermi velocity, the forward-scattering probability, and the dimensionless interaction strength.
The exact values of these parameters may depend on the transverse electric field and
the twist angle $\theta$.
For a typical experimental situation (see below), the frequencies of mTBG plasmons
should lie in the THz domain, which may be of interest for various applications. 
In this regard it is worthwhile to stress that mTBG is a natural plasmonic crystal.
It can be produced without advanced nanofabrication unlike artificial plasmonic crystals
with periodically-patterned gates~\mbox{\cite{Xiong2019, Bylinkin2019, Xiong2021}}.

We studied plasmons in two different regimes.
In the phase-coherent regime, $L \ll L_\varphi$,
where $L$ is the lattice constant of the network and $L_\varphi$
is the single-particle phase coherence length,
we employed the RPA.
In the phase-incoherent regime, $L_\varphi\ll L$, we used the PNM.
The PNM is similar to scattering models used to describe the single-particle band structure of the mTBG network~\cite{Efimkin2018, DeBeule2020};
however, it has two important differences. First, plasmon scattering at each network junction generically dissipates energy,
so the plasmon modes are damped everywhere except at a few high-symmetry points in the mini-Brillouin zone.
Second, the scattering at the junctions is governed by a full $6 \times 6$ matrix,
i.e., plasmons can scatter in any direction.
In contrast, in a related work,
plasmons were restricted to scatter only in certain directions, same as single particles~\cite{Brey2020}.
The omnidirectional plasmon scattering is possible because the Coulomb interaction
couples all electrons regardless of their valley index.
Indeed, the dielectric function of the system
depends on the sum of $\mathbf{P}^+$ and $\mathbf{P}^-$,
the polarization functions of the two valleys.
Therefore, the equation for the zero modes of the dielectric function 
necessarily involves a coupling between $\mathbf{P}^+$ and $\mathbf{P}^-$.

To realize the network of domain wall states in mTBG, two conditions must be satisfied: 
1) The moir\'e lattice constant $L$ must exceed the characteristic width of the domain walls $\ell\sim 5-10$ nm,
which requires that the twist angle $\theta\lesssim1^\circ$.
2)  The bandgap $\Delta_\text{TBG}$ must exceed the energy period $\varepsilon_L=2\pi \hbar v/L$   of the electronic states in the network. 
The largest achievable bandgaps in bilayer graphene are around 250meV~\mbox{\cite{Zhang2009}}, 
in which case  $\Delta_\text{TBG}\gtrsim\varepsilon_L$ requires that $\theta\lesssim1^\circ$. 
We therefore expect the network physics to become relevant at twist angles below about $1^\circ$. 
(However, to have a truly well-defined network we need $L\gg\ell$ and $\Delta_\text{TBG}\gg\varepsilon_L$, which requires $\theta\ll1^\circ$.)
Furthermore, in order to experimentally realize domain-wall plasmons in the network, it is necessary that the Fermi level resides in the band gap,
and that the frequency is low enough to avoid Landau damping associated with optical transitions to the bands in AB and BA regions. 
The former condition may be satisfied by using a top gate in addition to a back gate to independently control the displacement field and the Fermi level.
Of course, detecting plasmons in top-gated systems may be challenging.
Methods based on a transmission line could be a viable alternative
to near-field imaging~\cite{Gallagher2019, McIver2020}. 
To satisfy the latter condition, $\omega$ must be smaller than $\Delta_\mathrm{TBG} / 2 \hbar$,
which is $\approx 30\,\mathrm{THz}$ at $\Delta_\text{TBG}=250\,\mathrm{meV}$.
In comparison, a characteristic frequency scale of plasmons is $\omega_* = 2 \pi v / (K L)$.
Hence, with $v \sim 10^8\, \mathrm{cm}/\mathrm{s}$,
$L \sim 200\, \mathrm{nm}$ (corresponding to a twist angle of $\sim 0.07^{\circ}$), and $K \sim 0.5$, we have $\omega_* \sim 10\, \mathrm{THz}$. 
Therefore, 1D plasmons with wavelengths $\lambda_p < L$ can be realized.

In practice, at small twist angles there may be significant variations in the twist angle across the sample.
If the twist-angle disorder is weak, it may be accounted for by adding some additional broadening to the plasmon dispersion.  
Stronger disorder would wash out the plasmonic band structure, but long-wavelength plasmons 
with $\lambda_p\gg L$ may still be described by a continuum medium theory with a renormalized Drude weight.
We leave a detailed study of the disorder effects to future work.
\begin{figure}[t]
	\begin{center}
		\includegraphics[width=2.9in]{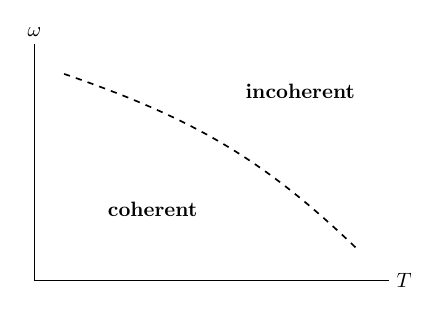}
	\end{center}
	\caption[Schematic diagram showing different regimes for the network versus temperature and interaction strength.]
	{Schematic regime diagram of the system versus temperature $T$ and frequency $\omega$.}
	\label{fig:phase_diagram}
\end{figure}

Other potential future directions for theoretical investigation include
exploring a crossover between the coherent and the incoherent regimes.
As discussed in Sec.~\ref{sec:Introduction2}, the coherence length $L_\varphi$ should decrease as the temperature or the frequency increases, 
so that a schematic regime diagram may look like as shown in Fig.~\ref{fig:phase_diagram}.
In the limits of weak forward scattering or weak deflection at the nodes,
this crossover can be studied by bosonization and perturbative renormalization-group techniques~\cite{Kane1992}.

The coherence length can be deduced from measuring Aharonov-Bohm oscillations
in magnetotransport experiments where mTBG is subject
to an external magnetic field~\cite{Rickhaus2018, Xu2019, DeBeule2020}.
In a fully coherent case the Aharonov-Bohm oscillations should evolve into
the quantum Hall effect with Hofstadter butterfly characteristics~\cite{Chou2020}.
It would be interesting to study the effect of the magnetic field on the plasmons;
in particular, on the localized eigenmodes that enclose a finite Aharonov-Bohm flux (Fig.~\ref{fig:localized_modes}).

Another intriguing possibility is to extend our model to near-commensurate twist angles, e.g., $\theta = 38.21^\circ$,
where TBG has also been predicted to host networks of topological domain wall states~\cite{Pal2019}.
Considering effects of elastic strain, which can induce quasi-1D channels in moir\'e systems~\cite{Sinner2023}, may also be worthwhile.
Narrow-band systems, such as TBG, are expected to possess undamped plasmons above the particle-hole continuum~\cite{Lewandowski2019}.
This may have implications for our model system as well.
Finally, it may be interesting to extend our theory to plasmons in artificial networks
of nanotubes or nanowires~\cite{Parage1998, Groep2012}.

\section{Acknowledgements}

We thank F. Guinea for pointing out to us the possibility of the Drude weight oscillations with doping and for discussion of the results.

\appendix

\section{Drude weight} \label{app:sp}

From Eq.~\eqref{eqn:drude} the average Drude weight is given by
\begin{equation}
\overline{D} = \frac{e^2}{\hbar v}\, \frac{N}{L} \sum_{l = 1}^3 \:
\int \frac{d^2 k}{(2 \pi)^2} {v}_{l x}^2(\mathbf k) ,
\label{eqn:average_Drude}
\end{equation}
where the integration is over the mini Brillouin zone of area $(2\pi)^2 / A_0$, $A_0 = \sqrt{3} L^2 / 2$ in the momentum space.
It is convenient to use $k_1 = \mathbf{k} \cdot \hat{\mathbf{l}}_1$ and $k_3 = \mathbf{k} \cdot \hat{\mathbf{l}}_3$ as two independent variables parametrizing other momenta.
For example,
$k_2 = \mathbf{k} \cdot \hat{\mathbf{l}}_2$ is given by $k_2 = -k_1 - k_3$.
The single-particle velocity in the $x$-direction can be written as
$v_{l x} = (\sqrt{3}\, / 2 \hbar) (\partial \varepsilon_l / \partial k_1)$.
Substituting this result into Eq.~\eqref{eqn:average_Drude}, we find
\begin{equation}
\overline{D} = \frac{\sqrt{3}}{2} \frac{e^2}{\hbar v}\, N L \sum_{l = 1}^3 \:
\int\limits_{-\pi / L}^{\pi / L}
\int\limits_{-\pi / L}^{\pi / L}
\frac{d k_1 d k_3}{(2 \pi \hbar)^2}
\left(\frac{\partial\varepsilon_l}{\partial k_1}\right)^2 .
\end{equation}
If $P_f = 1$, then $\varepsilon_l = \hbar v (k_l + \pi / L)$
and $\overline{D} = D_\mathrm{max}$ [Eq.~\eqref{eqn:D_max}].
For a weak deflection, $P_d \ll 1$,
we can use Eq.~\eqref{eqn:single-particle_bands_small_deflection}
to calculate the corrections $\Delta \overline{D}_{l l'}$ to $\overline{D}$. In particular,
\begin{align}
\Delta\overline{D}_{13} &= \frac{\sqrt{3}}{2} \frac{e^2}{\hbar v}\, N L
\int\limits_{-\pi / L}^{\pi / L}
\int\limits_{-\pi / L}^{\pi / L}
\frac{d k_1 d k_3}{(2 \pi \hbar)^2} \left[\sum_{\nu = \pm} \left(\frac{\partial\varepsilon_{13, \nu}}{\partial k_1}\right)^2 - (\hbar v)^2\right]
\nonumber \\
&= -D_\mathrm{max}
\int\limits_{-\pi / L}^{\pi / L}
\int\limits_{-\pi / L}^{\pi / L}
\frac{d k_1 d k_3}{(2 \pi)^2} \frac{L^2P_d}{(k_1 - k_3)^2 L^2 + 4 P_d}
\nonumber \\
&\simeq -D_\mathrm{max}
\int\limits_{-\infty}^\infty
\frac{d k}{2 \pi} \frac{LP_d}{k^2 L^2 + 4 P_d}
= -\frac14 D_\mathrm{max} \sqrt{P_d}.
\end{align}
Similarly, we find $\Delta\overline{D}_{23} = \Delta\overline{D}_{13}$ and $\Delta\overline{D}_{12} = 4 \Delta\overline{D}_{13}$.
Therefore,
$\overline{D} = D_\mathrm{max} + \Delta\overline{D}_{12} + \Delta\overline{D}_{13} + \Delta\overline{D}_{23}$ is given by Eq.~\eqref{eqn:D_avg}.

\section{Plasmon scattering at long wavelength}
\label{app:scattering}

Here we derive the leading-order correction to the formula
\begin{equation}
	e^{2 i \delta_m^{(0)}} = \frac{Y_* - Y_m}{Y_* + Y_m}
\end{equation}
[same as Eq.~\eqref{eqn:S_analytical}] for the plasmon scattering phase shifts $\delta_m$ 
for the case of a finite-range interaction.
We look for the solution of Eqs.~\eqref{eqn:F_m}, \eqref{eqn:bc_m},
and \eqref{eqn:bc_m_far} that is a superposition of incident and scattered waves:
\begin{align}
F(r) &= e^{-i q_p r} + F_{s}(r),
\label{eqn:V_s}\\
F_{s}(r) &\simeq -e^{2 i \delta_m} e^{i q_p r},
\quad r \to \infty\,.
\label{eqn:V_s_asym}
\end{align}
We treat the difference $\Delta U_m(r, s) \equiv U_m(r,s) - \tilde{U}(q_p) \delta(r - s)$
as a perturbation.
\begin{figure}[t]
	\begin{center}
		\includegraphics[width=2.75in]{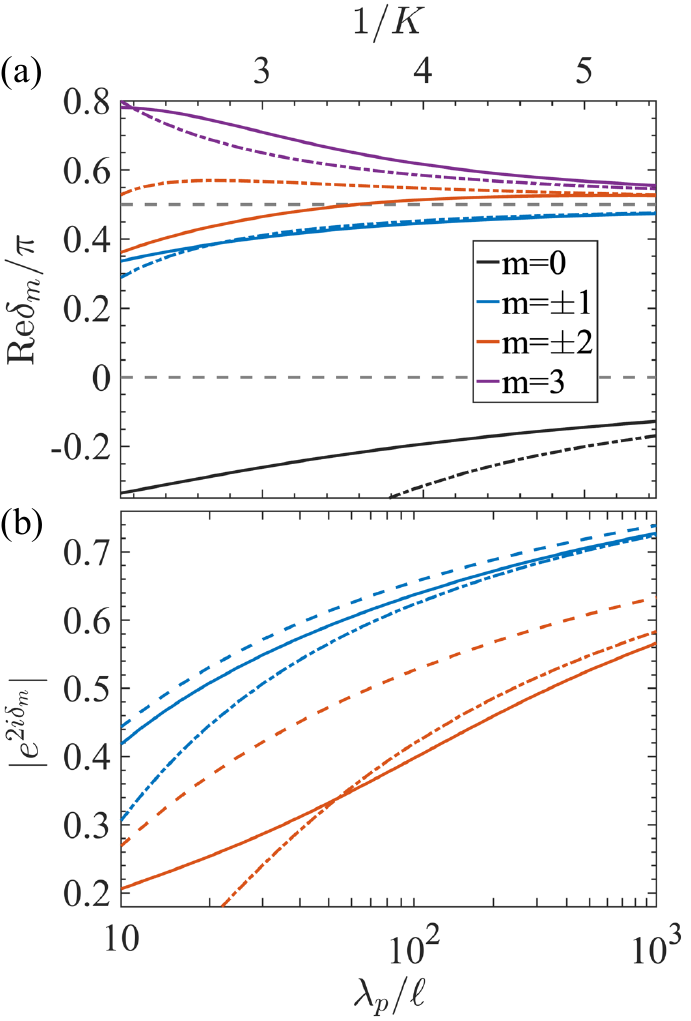}
	\end{center}
	\caption[Numerical results for scattering phase shifts.]
	{Scattering phase shifts for unscreened interaction versus the plasmon wavelength $\lambda_p$.
		The solid and the dashed lines are the same as in Fig.~\ref{fig:six_channel}.
		The dash-dotted lines include the first-order correction [Eq.~\eqref{eqn:delta_unscreened_corrections}] to the dashed lines.
		(a) Real parts of the phase shifts in different angular momentum channels.
		(b) Magnitude of the eigenvalues $e^{2 i \delta_m}$ of the $\mathbf{S}$ matrix for  $m = 1, 2$.
		\label{fig:six_channel_corrections}
	}
\end{figure}
Introducing the Green's function
\begin{equation}
	G_m^{(0)}(r,s) = -\frac{i}{2} q_p \left[ e^{i q_p |r - s|}
	- e^{2 i \delta_m^{(0)}} e^{i q_p (r + s)} \right] ,
	\label{eqn:G0}
\end{equation}
we obtain
\begin{align}
\begin{split}
F_{s}(r) =& -e^{2 i \delta_m^{(0)}} e^{i q_p r}
 - \frac{N v}{\pi \hbar \omega^2} \int\limits_0^\infty d s\, G_m^{(0)}(r, s)
 \\
 \mbox{} &\times 
  \frac{d}{d s} \int\limits_0^\infty d u\,  \Delta U_m(s, u)
\left[-i q_p e^{-i q_p u} + F_{s}'(u) \right] 
\end{split}
\label{eqn:V_m2}\\
\begin{split}
	\mbox{} =& -e^{2 i \delta_m^{(0)}} e^{i q_p r} + \frac{N v}{\pi \hbar \omega^2}\,
	\int\limits_0^\infty d s  \frac{\partial}{\partial s}  G_m^{(0)}(r, s)
	\\
	\mbox{} &\times	\int\limits_0^\infty d u\,
	\Delta U_m(s, u) 
	\left[-i q_p e^{-i q_p u} + F_{s}'(u)\right] .
\end{split}
\label{eqn:V_m3}
\end{align}
The last equation was derived using the integration by parts over $s$.
The $s = 0$ boundary term vanishes because $\Delta U_m(0, u) = 0$ for $m \neq 0$ and $G_m^{(0)}(r, 0) = 0$ for $m = 0$.
Substituting Eq.~\eqref{eqn:V_s_asym} for $F_{s}(u)$, we find the leading-order correction to the scattered field:
\begin{equation}
\begin{split}
F_{s}^{(1)}(r) =& -i q_p
\int\limits_0^\infty d s \int\limits_0^\infty d u\,
\left(e^{-i q_p u} + e^{2 i \delta_m^{(0)}} e^{i q_p u}\right)
\\
\mbox{} &\times
\frac{N v}{\pi \hbar \omega^2}\,
\Delta U_m(s, u) \frac{\partial}{\partial s}  G_m^{(0)}(r, s) .
\end{split}
\end{equation}
The large-$r$ asymptotic form of $F_{s}^{(1)}(r)$ is
\begin{equation}
\begin{split}
F_{s}^{(1)}(r) \simeq&  i q_p e^{i q_p r}
\int\limits_0^\infty d s \int\limits_0^\infty d u\,
\left(e^{-i q_p u} + e^{2 i \delta_m^{(0)}} e^{i q_p u}\right)
 \\
\mbox{} &\times
\frac{N}{2\pi \hbar v}\, K^2(q_p)
\Delta U_m(s, u)
 \left(e^{-i q_p s} + e^{2 i \delta_m^{(0)}} e^{i q_p s}\right) .
\end{split}
\label{eqn:V_s1}
\end{equation}
Comparing Eqs.~\eqref{eqn:V_s1} and \eqref{eqn:V_s_asym},
we obtain the correction to the phase shift
\begin{equation}
\begin{split}
\delta_m^{(1)} =& -q_p 
\int\limits_0^\infty d r
 \cos \left(q_p r + \delta_m^{(0)} \right)
 \int\limits_0^\infty d s
 \cos \left(q_p s + \delta_m^{(0)} \right)
\\
\mbox{} &\times
\frac{N}{\pi \hbar v}\, K^2(q_p)  \Delta U_m(r, s) .
\end{split}
\end{equation}
To evaluate the integrals,
we split $\Delta U_m(r, s)$ into one intra-lead term and five inter-lead terms
\begin{equation}
\begin{split}
\Delta U_m(r, s) =& [U(r - s) - \tilde{U}(q_p, 0) \delta(r - s)]
\\
 \mbox{} &+ \sum_{a = \pm 1, \pm 2, 3} e^{i \pi m a / 3}
 U\left(\sqrt{r^2 + s^2 - 2 r s \cos \frac{\pi a}{3}} \:\right) .
\end{split}
\label{eqn:Delta_U_m}
\end{equation}
In each of the resultant five integrals we change variables from $r$ and $s$ to
$u$ and $\theta$ such that
\begin{gather}
r = \frac{u}{\sin \beta} \cos \left(\theta - \frac{\beta}{2}\right),
\qquad
s = \frac{u}{\sin \beta} \cos \left(\theta + \frac{\beta}{2}\right),
\label{eqn:r_s_from_u}\\
u = \sqrt{r^2 + s^2 - 2 r s \cos \beta} ,
\qquad
\beta = \frac{\pi}{3} |a|,
\label{eqn:u}\\
-\frac{\pi - \beta}{2} < \theta < \frac{\pi - \beta}{2} .
\label{eqn:theta_range}
\end{gather}
The Jacobian of this transformation is
\begin{equation}
\left| \frac{\partial (r, s)}{\partial (u, \theta)} \right| = \frac{u}{\sin \beta} .
\label{eqn:Jacobian}
\end{equation}
For the case of screened interactions with $d_g \ll \lambda_p$,
each integral is proportional to
\begin{equation}
\frac{\pi - \beta}{\sin \beta} \int\limits_0^\infty d u\, U(u) u =
2\, \frac{\pi - \beta}{\sin \beta} \, \frac{e^2}{\kappa} d_g .
\label{eqn:uU_integral}
\end{equation}
\begin{widetext}
Altogether, we obtain
\begin{equation}
	%\begin{split}
		\delta_m^{(1)} = -\frac{2}{\pi}\, \alpha N K^2(q_p) q_p d_g
		\cos^2 \delta_m^{(0)}
		%\\
		%\mbox{} &\times
		\left[-1
		+ \frac{8\pi}{3\sqrt{3}} \cos \frac{\pi m}{3}
		+ \frac{4\pi}{3\sqrt{3}} \cos \frac{2\pi m}{3}
		+ (-1)^m \right] .
	%\end{split}
\label{eqn:delta_correction}
\end{equation}
This first-order perturbation theory formula should be valid if $(\alpha N/\pi)K^2 \ll 1$ in addition to $q_pd_g\ll1$.
For unscreened interactions, we find
\begin{equation}
\begin{split}
\delta_m^{(1)} =& -\frac{1}{2 \pi}\, \alpha N K^2(q_p) \left\{ \pi
\left[ -\frac12 \cos 2\delta_m^{(0)}
 + 2\cos \frac{\pi m}{3}
 + \frac{2}{\sqrt{3}} \cos \frac{2\pi m}{3}
 + \frac12 (-1)^m \right] \right. \\
& - \left. \sin 2\delta_m^{(0)}
\left[ \frac{4}{\sqrt{3}} \ln(2 + \sqrt{3})
\cos \frac{\pi m}{3} + 2\ln 3 \, \cos \frac{2 \pi m}{3}
 + (-1)^m \right] \right\}.
\label{eqn:delta_unscreened_corrections}
\end{split}
\end{equation}
For $q_p\ell\ll 1$, this is proportional to $1/\mathcal L$, where $\mathcal L = \ln(A / q_p \ell) \gg 1$ and $A = 0.561$ 
[cf. Eq.~\eqref{eqn:Fourier_kernel}].
In practice, it is difficult to make parameter $\mathcal{L}$ truly large.
For example, if $\lambda_p / \ell = 300$, then $\mathcal{L} = 3.3$.
Nevertheless, as shown in Fig.~\ref{fig:six_channel_corrections},
including these corrections improves the agreement between analytical and numerical results for $\delta_m$ at such $\lambda_p$.
\end{widetext}

%%%%%%%%%%%%%%%%%%%%%%%%%%%%%%%%%%%%%%%%%%%%%%%%%%%%%%%%%%%%%%%%%%
\section{Dielectric function of the local PNM}
\label{app:local}

Here we derive the dielectric function
given in Eqs.~\eqref{eqn:epsilon_local}--\eqref{eqn:B}.
In the local approximation, we replace the interaction kernel $U(x)$ in the equations of motion by $\tilde U(q_p)\delta(x)$,
where $q_p$ is the plasmon momentum at frequency $\omega$.
At a given crystal momentum $\mathbf q$,
let $n_a(x)$ and $V_{\mathrm{ext},a}(x)$ be respectively the density and external potential at position $x$ along link $a$.
From Eqs.~\eqref{eqn:continuity}-\eqref{eqn:Phi},
\begin{equation}
n_a(x)=\sum_{a'}\int\limits_0^L d x' P_{aa'}(x,x') e V_{\mathrm{ext}, a'}(x'),
\end{equation}
with
\begin{align}
	P_{aa'}(x,x') =& -\frac{N}{\pi\hbar v} K^2 \left[ \frac{i}{2}q_pe^{iq_p|x-x'|} + \delta(x-x') \right]\delta_{aa'} \nonumber \\
	&+ p_{aa',+}e^{iq_px} + p_{aa',-}e^{iq_p(L-x)}. \label{eqn:n_induced2}
\end{align}
The first term is the result for an infinite wire
and the two additional terms are due to scattering.
Throughout this Appendix, $K = K(q_p)$.
To find the coefficients $p_{aa',+}$ and $p_{aa',-}$
we impose boundary conditions at the network nodes through the scattering matrix $\mathbf S$.
Define six-component vectors $\mathbf n_\mathrm{in}$ and $\mathbf n_\mathrm{out}$ which
are respectively the incoming and outgoing plasmon amplitudes at a node.
Including the appropriate Bloch phases factors, they have components
\begin{align}
	n_{\mathrm{in},a+} &= e^{i(q_p-q_a)L} \left[p_{aa',+}-\frac{iq_p}{2} \frac{N}{\pi\hbar v}K^2 e^{-iq_px'}\right], \\
	n_{\mathrm{in},a-} &= \left[p_{aa',-} e^{iq_pL}-\frac{iq_p}{2} \frac{N}{\pi\hbar v}K^2 e^{iq_px'}\right], \\
	n_{\mathrm{out},a+} &= p_{aa',-} e^{-i q_aL}, \\
	n_{\mathrm{out},a-} &= p_{aa',+},
\end{align}
see Fig.~\ref{fig:junction_circuit} for the diagram explaining the link labels.
These amplitudes are related by $\mathbf{n}_\mathrm{out} = \mathbf{S}\,\mathbf{n}_\mathrm{in}$,
the solution of which yields
\begin{align}
	p_{aa',+} &= -\frac{iq_p}{2} \frac{N}{\pi\hbar v} K^2 \bigg[B_{a+, a'+} e^{-iq_px'} + B_{a+, a'-} e^{iq_p(x'-L)}\bigg],
	\label{eqn:n_a+}\\
	p_{aa',-} &= -\frac{iq_p}{2} \frac{N}{\pi\hbar v} K^2 \bigg[B_{a-, a'+} e^{-iq_px'} + B_{a-, a'-} e^{iq_p(x'-L)}\bigg],
	\label{eqn:n_a-}
\end{align}
where the matrix $\mathbf{B}$ is defined in Eq.~\eqref{eqn:B}.
Substituting Eqs.~\eqref{eqn:n_a+} and \eqref{eqn:n_a-} into Eq.~\eqref{eqn:n_induced2},
we find the real-space polarization function $P_{a a'}(x, x')$.
With $V_{\text{ext},a}(x) = \delta_{aa'}\delta(x-x')$, the induced potential is given by
\begin{equation}
\phi_{a a'}(x, x') = \tilde U(q_p)P_{a a'}(x, x') \,,
\end{equation}
which yields Eq.~\eqref{eqn:P_local_r}.
From this,
we find the  induced potential and inverse dielectric function in the 1D Fourier basis,
respectively Eqs.~\eqref{eqn:P_local} and \eqref{eqn:epsilon_local}.

\section{Non-local PNM} \label{app:non-local}

In this Appendix, we derive the equations of motion for the non-local PNM
and the resulting polarization function.
Using the notation introduced in Sec.~\ref{sec:RPA} of the main text,
the electron density, current, and external potential at a point a distance $x$ along the $a\,$th link in the unit cell specified by $\mathbf{R}$
are denoted by $n_{a, \mathbf{R}}(x)$, $j_{a, \mathbf{R}}(x)$, and $\Phi_{\mathrm{ext}.a, \mathbf{R}}(x)$, respectively.
The equations of motion for the PNM are
\begin{widetext}
\begin{gather}
-i\omega en_{a,\mathbf R}(x) + j_{a,\mathbf R}'(x) = \delta(x)\frac{\gamma Le}{3}\Big[ n_{a+1,\mathbf R}(0)+n_{a+1,\mathbf R-\mathbf l_{a+1}}(L) %\nonumber \\
-2n_{a,\mathbf R}(0)-2n_{a,\mathbf R-\mathbf l_a}(L)+n_{a-1,\mathbf R}(0)+n_{a-1,\mathbf R-\mathbf l_{a-1}}(L) \Big],
\label{eqn:density} \\
-i\omega j_{a,\mathbf R}(x) + \frac{Ne^2v}{\pi\hbar}V_{a,\mathbf R}'(x) =   \delta(x)\frac{\gamma L}{3}\Big[ j_{a+1,\mathbf R}(0)+j_{a+1,\mathbf R-\mathbf l_{a+1}}(L) %\nonumber \\
-2j_{a,\mathbf R}(0)-2j_{a,\mathbf R-\mathbf l_a}(L)+j_{a-1,\mathbf R}(0) %\nonumber \\
+j_{a-1,\mathbf R-\mathbf l_{a-1}}(L) \Big],
\label{eqn:current} \\
eV_{a,\mathbf R}(x) = \frac{\pi\hbar v}{N}n_{a,\mathbf R}(x) %\nonumber \\
+ \sum_{a',\mathbf R'}\int\limits_0^Ldx' U(\mathbf R+\hat{\mathbf l}_ax-\mathbf R'-\hat{\mathbf l}_{a'}x')n_{a',\mathbf R'}(x') %\nonumber \\
+ e\Phi_{\mathrm{ext},a,\mathbf R}(x).
\label{eqn:voltage}
\end{gather}
\end{widetext}
Here, as before, all the link indices $a$, $a'$, \textit{etc}., are modulo $3$.
The terms on the right-hand-side of Eqs.~\eqref{eqn:density} and \eqref{eqn:current} enforce the boundary conditions
given in Eq.~\eqref{eqn:F_bc_multi} of the main text.
Note that these six terms are the densities and currents on the six links meeting at a network node.
At a given crystal momentum $\mathbf{q}$, we expand the electron density, current, and external potential along each link as
\begin{align}
n_{a,\mathbf R}(x) &= e^{i\mathbf q\cdot\mathbf R}\sum_{p=-\infty}^\infty n_{ap}e^{iq_{ap}x},
\label{eqn:density_Fourier} \\
j_{a,\mathbf R}(x) &= e^{i\mathbf q\cdot\mathbf R}\sum_{p=-\infty}^\infty j_{ap}e^{iq_{ap}x},
\label{eqn:current_Fourier} \\
\Phi_{\mathrm{ext},a,\mathbf R}(x) &= e^{i\mathbf q\cdot\mathbf R}\sum_{p=-\infty}^\infty \Phi_{\mathrm{ext},ap}e^{iq_{ap}x} \,.
\label{eqn:potential_Fourier}
\end{align}
Substituting Eqs.~\eqref{eqn:density_Fourier}-\eqref{eqn:potential_Fourier} into Eqs.~\eqref{eqn:density}-\eqref{eqn:voltage}, we find
\begin{align}
\omega en_{ap} - q_{ap}j_{ap} %\nonumber \\
=& \frac{i\gamma}{3}\sum_{p'}(n_{a+1,p'}-2n_{ap'}+n_{a-1,p'}),
\label{eqn:density2} \\
\omega j_{ap} - ev^2q_{ap}n_{ap} %\nonumber \\
=& \frac{i\gamma}{3}\sum_{p'}(j_{a+1,p'}-2j_{ap'}+j_{a-1,p'}) \nonumber \\
&+ \frac{Ne^2v}{\pi\hbar}q_{ap}\Phi_{ap},
\label{eqn:current2}
\end{align}
where
\begin{equation}
e\Phi_{ap} = \sum_{a'p'}\tilde U_{ap,a'p'}(\mathbf q)n_{a'p'}+e\Phi_{\mathrm{ext},ap},
\end{equation}
with $\tilde U_{ap,a'p'}(\mathbf q)$ defined in Eq.~\eqref{eqn:U1}.
To find the polarization function, we solve Eqs.~\eqref{eqn:density2} and \eqref{eqn:current2} for density $n_{ap}$ in terms of the potential $\Phi_{ap}$.
In terms of the chiral currents $j^{\nu}_{a p} = j_{a p} + \nu e n_{a p}$, where $\nu = \pm$, these equations become
\begin{equation}
\begin{split}
	(\omega - \nu v q_{a p}) j^{\nu}_{a p} =& \frac{i\gamma}{3} \sum_{p'} \left(
	 j^{\nu}_{a+1, p'} - 2 j^{\nu}_{a p'} + j^{\nu}_{a-1, p'}\right) \\
	 \mbox{} &+ \frac{N e^2}{\pi \hbar} v q_{a p} \Phi_{a p} \,.
\end{split}
\label{eqn:3}
\end{equation}
Now divide through by $\omega - \nu v q_{a p}$ and sum over $p$ to obtain
\begin{equation}
\begin{split}
	\sum_p j^{\nu}_{a p} =& \frac{1}{3} \lambda^{\nu}_{a}\,
	\sum_p \left(j^{\nu}_{a+1, p} - 2 j^{\nu}_{a p} + j^{\nu}_{a-1, p}\right) \\
	&+ \sum_{p} \frac{v q_{a p}}{\omega - \nu v q_{a p}}
	 \frac{N e^2}{\pi \hbar} \Phi_{a p} \,,
\end{split}
\label{eqn:4}
\end{equation}
with $\lambda^{\nu}_{a}$ given by Eq.~\eqref{eqn:lambda_s_a}.
Solving this system of three equations for the variables $\sum_p j^{\nu}_{a p}$,
we find
\begin{equation}
	\sum_p \left(j^{\nu}_{a+1, p} - 2 j^{\nu}_{a p} + j^{\nu}_{a-1, p}\right)
	= \frac{3 N e^2}{\pi \hbar} \sum_{a'p'} \mathcal{G}^{\nu}_{a a'} \frac{v q_{a' p'}}{\omega - \nu v q_{a' p'}}
	\Phi_{a'p'} ,
	\label{eqn:6}
\end{equation}
where $\mathcal{G}^{\nu}_{a a'}$ are the elements of the matrix
defined by Eq.~\eqref{eqn:G_matrix}.
Therefore,
\begin{equation}
\begin{split}
	j^{\nu}_{a p} =& \frac{N e^2}{\pi \hbar} \frac{1}{\omega - \nu v q_{a p}}
	 \Bigg[ v q_{a p} \Phi_{a p} \\
	  \mbox{} &+ i \gamma \sum_{a'p'} \mathcal{G}^{\nu}_{a a'}(\mathbf{q}, \omega) \frac{v q_{a' p'}}{\omega - \nu v q_{a' p'}} \Phi_{a'p'}
	  	\Bigg] .
\end{split}
\label{eqn:8}
\end{equation}
This yields the polarization function of valley $\nu$ given by Eq.~\eqref{eqn:polarization3}.

For the effective dielectric function [Eq.~\eqref{eqn:epsilon_00}] we obtain
\begin{align}
\epsilon_{\mathrm{EMT}}(\mathbf{q}, \omega) \simeq& 1 - \frac{D_\mathrm{max}}{\pi e^2}\,\tilde{U}(\mathbf{q})
q^2
\nonumber \\
	 \mbox{} &\times \Bigg[
	 \frac{2(\omega + i \gamma) + C v q}
	      {4 \omega (\omega + i \gamma)^2 - v^2 q^2 (3 \omega + 2 i \gamma) - C v^3 q^3}
\nonumber \\
\mbox{} &+ \left(C \to -C\right) \Bigg],
\label{eqn:epsilon_eff_expanded}
\end{align}
where $C = \cos 3 \theta$.
To get this formula, we used the approximation
\begin{equation}
 \lambda_a^\nu \simeq \frac{i\gamma}{\omega - \nu v q_a}\,,
 \qquad  \omega, v q \ll \frac{v}{L}\,.
\end{equation}
From Eq.~\eqref{eqn:epsilon_eff_expanded} we can derive Eqs.~\eqref{eqn:epsilon_EMT}, \eqref{eqn:slow_diffusion_pole}, and \eqref{eqn:Z} in appropriate limits
and also show that Eq.~\eqref{eqn:2d_plasmon_short-range} becomes valid if $\gamma / v \ll q \ll 1 / L$.

\bibliography{plasmons_mTBG_network}
\end{document}